%% file: isec.tex
\documentclass[nonacm, sigconf]{acmart}

\renewcommand\footnotetextcopyrightpermission[1]{} 
\usepackage{graphicx, subfig}
\usepackage[inline]{enumitem}
\usepackage{multirow}
\usepackage{xspace}

\usepackage{hyperref}
\hypersetup{
  colorlinks=true,      
  linkcolor=blue,       
  citecolor=magenta,    
  filecolor=cyan,       
  urlcolor=red          
}

\date{}

\def\ie{{i.e.}}
\def\eg{{e.g.}}

\newenvironment{denseitemize}{
\begin{itemize}[topsep=2pt, partopsep=0pt, leftmargin=1em]
  \setlength{\itemsep}{2pt}
  \setlength{\parskip}{0pt}
  \setlength{\parsep}{0pt}
}{\end{itemize}}

\newenvironment{denseenum}{
\begin{enumerate}[topsep=2pt, partopsep=0pt, leftmargin=1.5em]
  \setlength{\itemsep}{2pt}
  \setlength{\parskip}{0pt}
  \setlength{\parsep}{0pt}
}{\end{enumerate}}

\newcommand{\rev}[1]{\textcolor{black}{#1}}

\newcommand{\name}{Hydra\xspace}
\newcommand{\copysets}{CodingSets\xspace}
\newcommand{\is}{Infiniswap}
\newcommand{\host}{Resilience Manager\xspace}
\newcommand{\daemon}{Resource Monitor}
\newcommand{\Prob}[1]{\P\!\left[ {#1} \right]}
\renewcommand{\P}{\mathbb{P}}

\begin{document}
\title{\rev{\rm \bf{{\name} : Resilient and Highly Available Remote Memory}}}

\author{Youngmoon Lee$^{*1}$, Hasan Al Maruf$^{*1}$, Mosharaf Chowdhury$^1$,  Asaf Cidon$^2$, Kang G. Shin$^1$ }

\affiliation{\institution{University of Michigan$^1$ \hspace{0.4em} Columbia University$^2$}
}


\renewcommand{\shortauthors}{}
\input{abstract}
\maketitle

\thispagestyle{empty}
\def\thefootnote{*}\footnotetext{These authors contributed equally to this work}\def\thefootnote{\arabic{footnote}}

\input{intro}
\input{motivation}
\input{overview}
\input{design}

\input{analysis}

\input{implementation}
\input{evaluation}
\input{related}
\input{outro}
\input{acknowledgments}
\label{EndOfPaper}


\bibliographystyle{ACM-Reference-Format}
\bibliography{isec}
\end{document}

%% file: abstract.tex
\begin{abstract}

We present \name, a low-latency, low-overhead, and highly available resilience mechanism for remote memory. 
{\name} can access erasure-coded remote memory within a single-digit $\mu$s read/write latency, significantly improving the performance-efficiency tradeoff over the state-of-the-art -- it performs similar to in-memory replication with 1.6$\times$ lower memory overhead.
We also propose {\copysets}, a novel coding group placement algorithm for erasure-coded data, that provides load balancing while reducing the probability of data loss under correlated failures by an order of magnitude. 
With {\name}, even when only 50\% memory is local, unmodified memory-intensive applications achieve performance close to that of the fully in-memory case in the presence of remote failures and outperforms the state-of-the-art solutions by up to $4.35\times$.
\end{abstract}

%% file: intro.tex
\section{Introduction}



Modern datacenters are embracing a paradigm shift toward disaggregation, where each resource is decoupled and connected through a high-speed network fabric~\cite{intel-rack, fb-rack, open-rack, legoos, far-memory, dredbox, firebox, mem-ext, hp-server-consolidation, hp-machine}.
In such disaggregated datacenters, each server node is specialized for specific purposes -- some are specialized for computing, while others for memory, storage, and so on.
Memory, being the prime resource for high-performance services, is becoming an attractive target for disaggregation~\cite{mem-ext, mem-disagg-vmware, infiniswap, remote-regions, app-performance-disagg-dc, far-memory, relationaldb-rdma, hyper, kona}.

Recent remote-memory frameworks allow an unmodified application to access remote memory in an implicit manner via well-known abstractions such as distributed virtual file system (VFS) and distributed virtual memory manager (VMM)~\cite{remote-regions, infiniswap, app-performance-disagg-dc, leap, legoos, far-memory, disaggregated-dbms}. 
With the advent of RDMA, remote-memory solutions are now close to meeting the single-digit $\mu$s latency required to support acceptable application-level performance \cite{app-performance-disagg-dc, far-memory}. 
However, realizing remote memory for heterogeneous workloads running in a large-scale cluster faces considerable challenges \cite{mem-disagg-vmware, hotnets-disagg-failure} stemming from two root causes:

\begin{denseenum}
  \item \emph{Expanded failure domains:} As applications rely on memory across multiple machines in a remote-memory cluster, they become susceptible to a wide variety of failure scenarios. 
  Potential failures include independent and correlated failures of remote machines, evictions from and corruptions of remote memory, and network partitions.

  \item \emph{Tail at scale:} Applications also suffer from stragglers or late-arriving remote responses. 
  Stragglers can arise from many sources including latency variabilities in a large network due to congestion and background traffic \cite{tail-scale}.
\end{denseenum}
While one leads to catastrophic failures and the other manifests as service-level objective (SLO) violations, both are unacceptable in production \cite{far-memory, nines-not-enough}. 
Existing solutions take three primary approaches to address them: 
(i) local disk backup \cite{infiniswap, legoos},
(ii) remote in-memory replication \cite{markatos-remote, net-ramdisk, fasst, farm}, and 
(iii) remote in-memory erasure coding \cite{XORingEN, eccache, memec, cocytus} and compression \cite{far-memory}.
Unfortunately, they suffer from some combinations of the following problems.

\begin{figure}[!t]
	\centering
	\includegraphics[width=0.85\columnwidth]{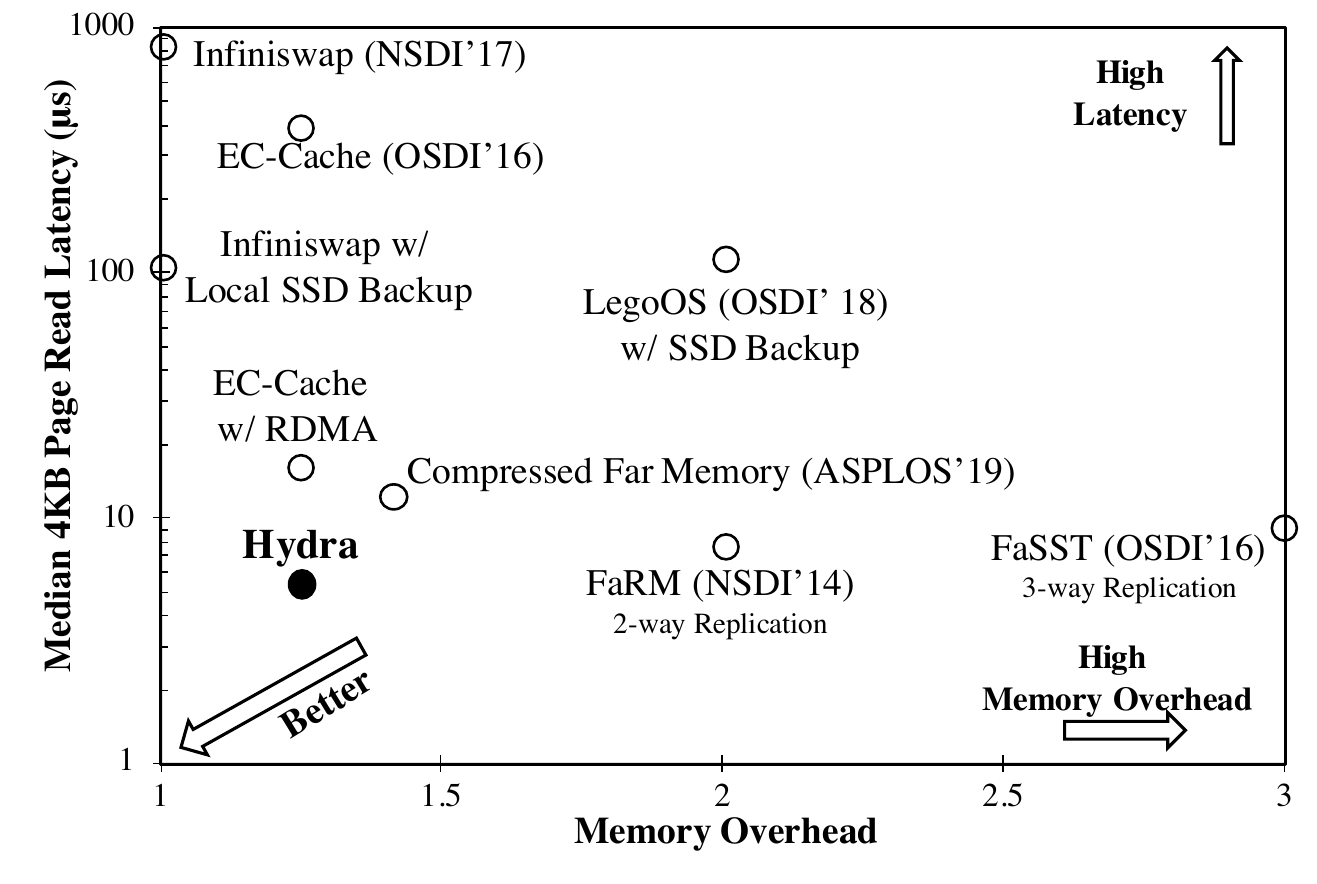}
	\caption{Performance-vs-efficiency tradeoff in the resilient cluster memory design space.
    Here, the Y-axis is in log scale.
    }
	\label{fig:latency-vs-storage}
\end{figure}

\textbf{High latency:} Disk backup has no additional memory overhead, but the access latency is intolerably high under any correlated failures. 
Systems that take the third approach do not meet the single-digit $\mu$s latency requirement of remote memory even when paired with RDMA (Figure~\ref{fig:latency-vs-storage}).

\textbf{High cost:} Replication has low latency, but it doubles memory consumption and network bandwidth requirements. 
Disk backup and replication represent the two extreme points in the performance-vs-efficiency tradeoff space 
(Figure~\ref{fig:latency-vs-storage}).
%

\textbf{Low availability:} All three approaches lose availability to low latency memory when even a very small number of servers become unavailable.
With the first approach, if a single server fails its data needs to be reconstituted from disk, which is a slow process.
In the second and third approach, when even a small number of servers (\eg, three) fail simultaneously, some users will lose access to data. 
This is due to the fact that replication and erasure coding assign replicas and coding groups to \emph{random} servers.
Random data placement is susceptible to data loss when a small number of servers fail at the same time~\cite{copyset,tiered} (Figure~\ref{fig:availability-vs-storage}).

In this paper, we present {\name}, a low-latency, low-overhead, and highly available resilience mechanism for remote memory to mitigate these problems. 
\rev{While erasure codes are known for reducing storage overhead and for better load balancing, it is challenging for remote memory with $\mu$s-scale access requirements (preferably, 3-5$\mu$s)~\cite{app-performance-disagg-dc}.
We demonstrate how to achieve resilient erasure-coded cluster memory with single-digit $\mu$s latency even under simultaneous failures at a reduced data amplification overhead.}

We explore the challenges and tradeoffs for resilient remote memory without sacrificing application-level performance or incurring high overhead in the presence of correlated failures (\S\ref{sec:motivation}). 
We also explore the trade-off between load balancing and high availability in the presence of simultaneous server failures.
Our solution, {\name}, is a configurable resilience mechanism that applies online erasure coding to individual remote memory pages while maintaining high availability (\S\ref{sec:arch}). 
{\name}'s carefully designed data path enables it to access remote memory pages within a single-digit $\mu$s median and tail latency (\S\ref{sec:design}). 
Furthermore, we develop \copysets, a novel coding group placement algorithm for erasure codes that provides load balancing while reducing the probability of data loss under correlated failures (\S\ref{sec:analysis}).

\rev{We develop {\name} as a drop-in resilience mechanism that can be applied to existing remote memory frameworks~\cite{remote-regions, infiniswap, leap, legoos, kona}.}
We integrate {\name} with the two major remote memory approaches widely embraced today: disaggregated VMM (used by Infiniswap \cite{infiniswap}, and Leap \cite{leap}) and disaggregated VFS (used by Remote Regions \cite{remote-regions}) (\S\ref{sec:implementation}). 
Our evaluation using production workloads shows that {\name} achieves the best of both worlds (\S\ref{sec:eval}).
{\name} closely matches the performance of replication-based resilience with 1.6$\times$ lower memory overhead with or without the presence of failures.
At the same time, it improves latency and throughput of the benchmark applications by up to 64.78$\times$ and 20.61$\times$, respectively, over SSD backup-based resilience with only 1.25$\times$ higher memory overhead. 
While providing resiliency, {\name} also 
 improves the application-level performance by up to $4.35\times$ over its counterparts.
\copysets reduces the probability of data loss under simultaneous server failures by about 10$\times$.
\rev{{\name} is available at \href{https://github.com/SymbioticLab/hydra}{https://github.com/SymbioticLab/hydra.}}

In this paper, we make the following contributions:
\begin{denseitemize}
  \item {\name} is the first in-memory erasure coding scheme that achieves single-digit $\mu$s tail memory access latency.
  
  \item Novel analysis of load balancing and availability trade-off for distributed erasure codes.
  
  \item \copysets is a new data placement scheme that balances availability and load balancing, while reducing probability of data loss by an order of magnitude during  failures.
\end{denseitemize}

%% file: motivation.tex
\section{Background and Motivation}
\label{sec:motivation}

\subsection{Remote Memory}
\label{subsec:model}

Remote memory exposes memory available in remote machines as a pool of memory shared by many machines. 
It is often implemented logically by leveraging stranded memory in remote machines via well-known abstractions, such as the file abstraction \cite{remote-regions}, remote memory paging \cite{infiniswap, app-performance-disagg-dc, swapping-infiniband, leap, kona}, and virtual memory management for distributed OS \cite{legoos}.
In the past, specialized memory appliances for physical memory disaggregation were proposed \cite{mem-ext, mem-ext-sys}. 

\begin{figure}[!t]
	\centering
	\includegraphics[width=0.85\columnwidth]{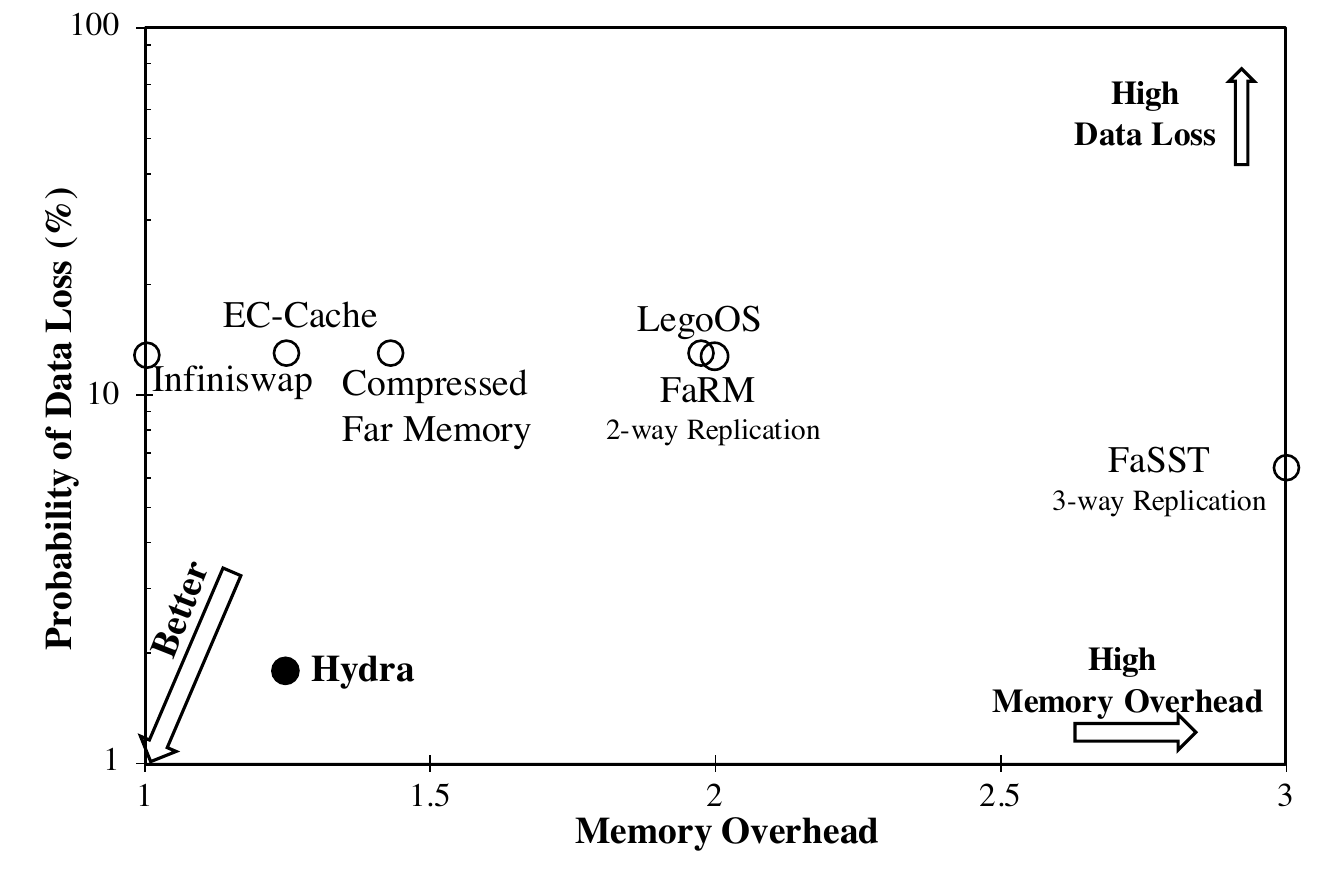}
	\caption{Availability-vs-efficiency tradeoff considering 1\% simultaneous server failures in a 1000-machine cluster.
    }
	\label{fig:availability-vs-storage}
\end{figure}

All existing remote-memory solutions use the 4KB page granularity.
While some applications use huge pages for performance enhancement \cite{ingens}, the Linux kernel still performs paging at the basic 4KB level by splitting individual huge pages because huge pages can result in high amplification for dirty data tracking \cite{pberry}.
%
Existing remote-memory systems use disk backup \cite{infiniswap, legoos} and in-memory replication \cite{markatos-remote, net-ramdisk} to provide availability during failures.

\begin{figure*}[!t]
	\centering    
	\subfloat[][Remote failure]{%
		\label{fig:disk-failure}
		\begin{minipage}{0.24\textwidth}
			\includegraphics[width=\textwidth]{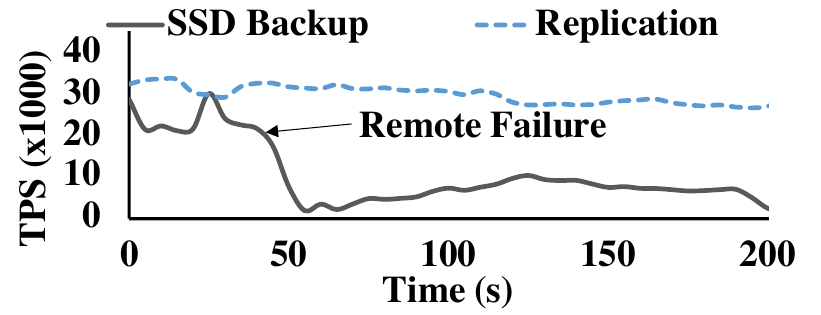}
		\end{minipage}
	}        
	\hfill
	\subfloat[][Background Network Load]{%
		\label{fig:disk-load}%
		\begin{minipage}{0.24\textwidth}
			\includegraphics[width=\textwidth]{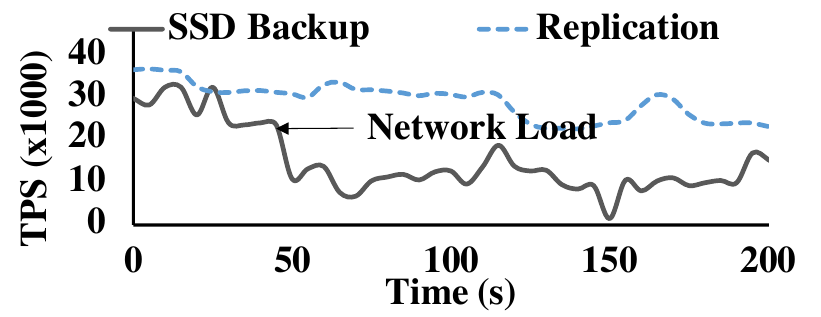}
		\end{minipage}
	}
	\hfill
	\subfloat[][Request burst]{
		\label{fig:disk-burst}
		\begin{minipage}{0.24\textwidth}
			\includegraphics[width=\textwidth]{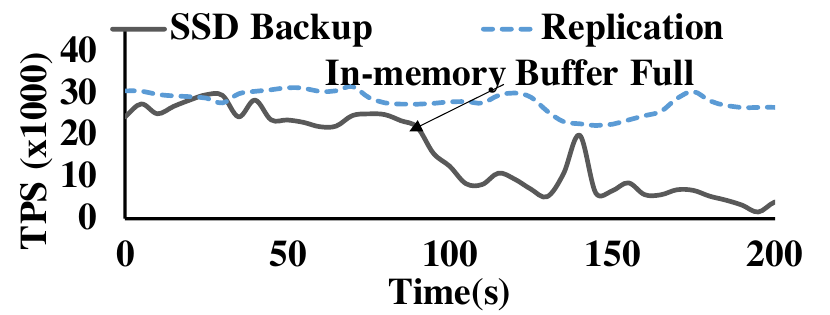}
		\end{minipage}
	}
	\subfloat[][Page Corruption]{%
	 	\label{fig:disk-corruption}%
	 	\begin{minipage}{0.24\textwidth}
	 		\includegraphics[width=\textwidth]{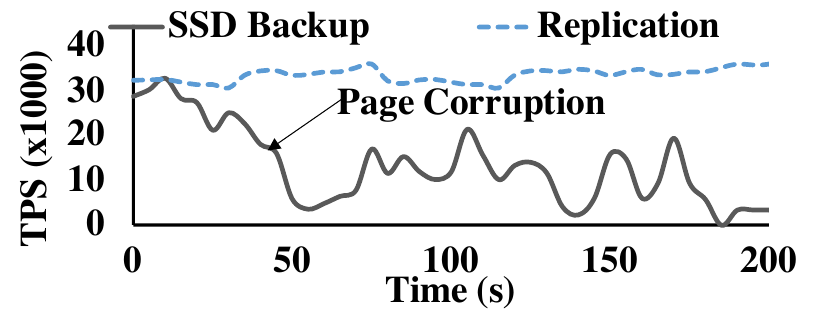}
	 	\end{minipage}
	 }    
	\hfill
	\caption{
	TPC-C throughput on VoltDB when 50\% of the working set fits in memory. Arrows point to uncertainty injection time.
	}
	\label{fig:motivation}
\end{figure*}

\subsection{Failures in Remote Memory}
\label{subsec:uncertainties}

The probability of failure or temporary unavailability is higher in a large remote-memory cluster, since memory is being accessed remotely.
To illustrate possible performance penalties in the presence of such unpredictable events, we consider a resilience solution from the existing literature \cite{infiniswap}, where each page is asynchronously backed up to a local SSD.
We run transaction processing benchmark TPC-C \cite{tpc-c} on an in-memory database system, VoltDB \cite{voltdb}.
We set VoltDB's available memory to 50\% of its peak memory to force remote paging for up to 50\% of its working set.

\paragraph{\textbf{1. Remote Failures and Evictions}}
Machine failures are the norm in large-scale clusters
where thousands of machines crash over a year due to a variety of reasons, including software and hardware failures~\cite{google-trace1, hadoop-availability, google-future, unavailability-study}.
Concurrent failures within a rack or network segments are quite common and typically occur dozens of times a year. 
Even cluster-wide power outage  is not uncommon -- occurs once or twice per year in a given data center.
For example, during a recent cluster-wide power outage in Google Cloud, around 23\% of the machines were unavailable for hours~\cite{google-outage}.

Without redundancy, applications relying on remote memory may fail when a remote machine fails or remote memory pages are evicted. 
As disk operations are significantly slower than the latency requirement of remote memory, disk-based fault-tolerance is far from being practical.
In the presence of a remote failure, VoltDB experiences almost 90\%  throughput loss (Figure~\ref{fig:disk-failure}); throughput recovery takes a long time after the failure happens.

\paragraph{\textbf{2. Background Network Load}}
Network load throughout a large cluster can experience significant fluctuations \cite{dc-traffic, tail-scale}, which can inflate RDMA latency and application-level stragglers, causing unpredictable performance issues \cite{frdma, justitia}.
In the presence of an induced bandwidth-intensive background load, VoltDB throughput drops by about 50\% (Figure~\ref{fig:disk-load}).

\paragraph{\textbf{3. Request Bursts}}
Applications can have bursty memory access patterns. 
Existing solutions maintain an in-memory buffer to absorb temporary bursts \cite{ramcloud, remote-regions, infiniswap}. 
However, as the buffer ties remote access latency to disk latency when it is full, the buffer can become the bottleneck when a workload experiences a prolonged burst.
While a page read from remote memory is still fast, backup page writes to the local disk become the bottleneck after the $100^{th}$ second in Figure~\ref{fig:disk-burst}. 
As a result, throughput drops by about 60\%. 

 \paragraph{4. Memory Corruption}
During remote memory access, if any one of the remote servers experiences a corruption, or if the memory gets corrupted over the network a memory corruption event will occur and disk access causes failure-like performance loss (Figure~\ref{fig:disk-corruption}).

\paragraph{\textbf{Performance vs. Efficiency Tradeoff for Resilience}}
In all of these scenarios, the obvious alternative -- in-memory 2$\times$ or 3$\times$ replication \cite{markatos-remote, net-ramdisk} -- is effective in mitigating a small-scale failure, such as the loss of a single server (Figure~\ref{fig:disk-failure}).
When an in-memory copy becomes unavailable, we can switch to an alternative. 
Unfortunately, replication incurs high memory overhead in proportion to the number of replicas. 
This defeats the purpose of remote memory.
Hedging requests to avoid stragglers \cite{tail-scale} in a replicated system doubles its bandwidth requirement as well.

This leads to an impasse: one has to either settle for high latency in the presence of a failure or incur high memory overhead.
Figure~\ref{fig:latency-vs-storage} depicts this performance-vs-efficiency tradeoff 
under failures and memory usage overhead to provide resilience. 
Beyond these two extremes, there are two primary alternatives to achieve high resilience with low overhead. 
The first is replicating pages to remote memory after compressing them (\eg, using zswap) \cite{far-memory}, which improves the tradeoff in both dimensions.
However, its latency can be more than 10$\mu$s when data is in remote memory. Especially, during resource scarcity, the presence of a prolonged burst in accessing remote compressed pages can even lead to  orders of magnitude higher latency due to the demand spike in both CPU and local DRAM consumption for decompression.
Besides, this has similar issues as replication such as latency inflation due to stragglers.

The alternative is erasure coding, which has recently made its way from disk-based storage to in-memory cluster caching to reduce storage overhead and improve load balancing \cite{ec-storage, codfs, ceph, cocytus, memec, eccache}.
Typically, an object is divided into $k$ \emph{data splits} and encoded to create $r$ equal-sized \emph{parity splits} ($k > r$), which are then distributed across $(k+r)$ failure domains.
Existing erasure-coded memory solutions deal with large objects (\eg, larger than 1 MB \cite{eccache}), where hundreds-of-$\mu$s latency of the TCP/IP stack can be ignored.
Simply replacing TCP with RDMA is not enough either. 
For example, the EC-Cache with RDMA (Figure~\ref{fig:latency-vs-storage}) provides a lower storage overhead than compression but with a latency around 20$\mu$s.

Last but not least, all of these approaches experience high unavailability in the presence of correlated failures \cite{copyset}.

\subsection{Challenges in Erasure-Coded Memory}
\label{subsec:challenge}

\paragraph{\textbf{High Latency}}
Individually erasure coding 4 KB pages that are already small lead to even smaller data chunks ($\frac{4}{k}$ KB), which contributes to the higher latency of erasure-coded remote memory over RDMA due to following primary reasons: 
\begin{denseenum}
	\item \textbf{\textit{Non-negligible coding overhead:}}
	When using erasure codes with on-disk data or over slower networks that have hundreds-of-$\mu$s latency, its 0.7$\mu$s encoding and 1.5$\mu$s decoding overheads can be ignored. 
	However, they become non-negligible when dealing with DRAM and RDMA.
	
	\item \textbf{\textit{Stragglers and errors:}}
	As erasure codes require $k$ splits before the original data can be constructed, any straggler can slow down a remote read. 
	To detect and correct an error, erasure codes require additional splits; an extra read adds another round-trip to double the overall read latency. 
	
	\item \textbf{\textit{Interruption overhead:}}
	Splitting data also increases the total number of RDMA operations for each request. 
	Any context switch in between can further add to the latency.
	
	\item \textbf{\textit{Data copy overhead:}}
	In a latency-sensitive system, additional data movement can limit the lowest possible latency. 
	During erasure coding, additional data copy into different buffers for data and parity splits can quickly add up.

\end{denseenum}

\paragraph{\textbf{Availability Under Simultaneous  Failures}}
Existing erasure coding schemes can handle a small-scale failure without interruptions.
However, when a relatively modest number of servers fail or become unavailable at the same time (\eg, due to a network partition or a correlated failure event),
they are highly susceptible to losing availability to some of the data.

This is due to the fact that existing erasure coding schemes generate coding groups on random sets of servers~\cite{eccache}.
In a coding scheme with $k$ data and $r$ parity splits, an individual coding group, will fail to decode the data if $r+1$ servers fail simultaneously. 
Now in a large cluster with $r+1$ failures, the probability that those $r+1$ servers will fail for a \emph{specific} coding group is low.
However, when coding groups are generated randomly (\ie, each one of them compromises a random set of $k+r$ servers), and there are a large number of coding groups per server, then the probability that those $r+1$ servers will affect \emph{any} coding group in the cluster is much higher.
Therefore, state-of-the-art erasure coding schemes, such as EC-Cache, will experience a very high probability of unavailability even when a very small number of servers fail simultaneously.


%% file: overview.tex
\section{{\name} Architecture}
\label{sec:arch}
\begin{figure}[!t]
	\centering
	\includegraphics[width=\columnwidth]{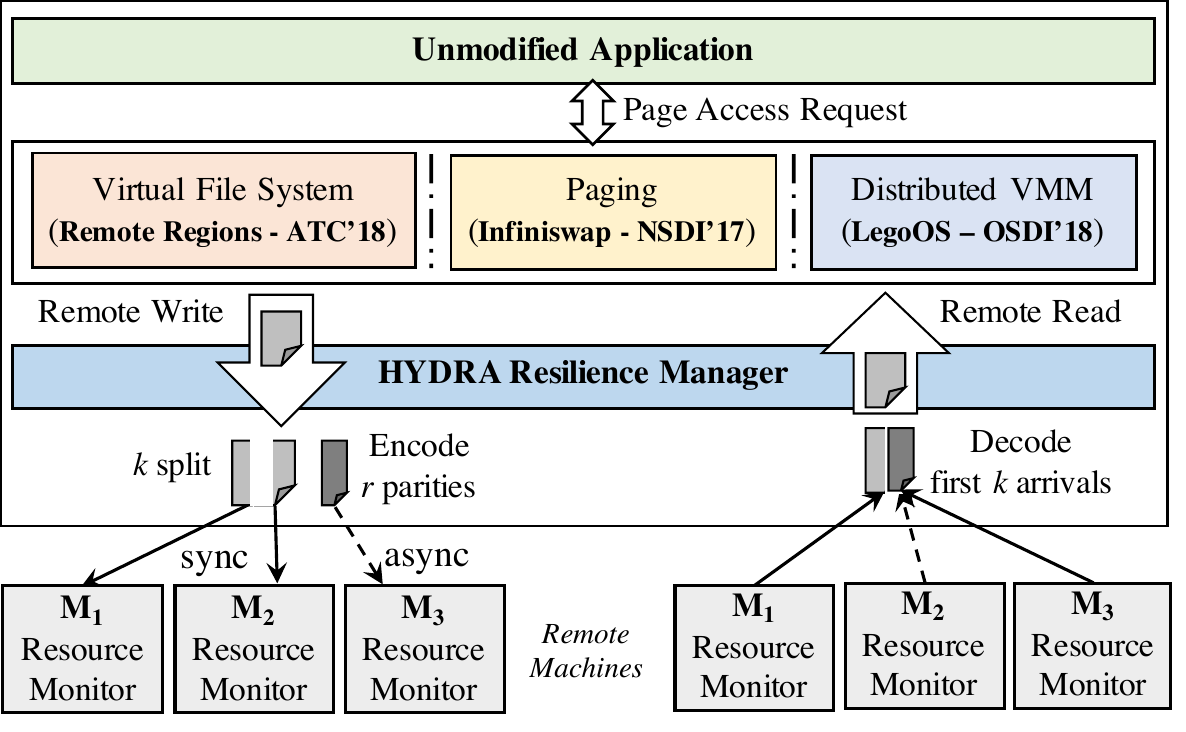}
	\caption{
	{\host} provides with resilient, erasure-coded remote memory abstraction.
	{\daemon} manages the remote memory pool.
	}
	\label{fig:overview}
\end{figure}

{\name} is an erasure-coded resilience mechanism for existing remote-memory techniques to provide better performance-efficiency tradeoff under remote failures while ensuring high availability under simultaneous failures. 
It has two main components (Figure~\ref{fig:overview}): 
(i) \textbf{ {\host}} coordinates erasure-coded resilience operations during remote read/write;
(ii) \textbf{ {\daemon}} handles the memory management in a remote machine.
Both can be present in every machine and work together without central coordination.

\subsection{{\host}}
{\name} {\host} provides remote memory abstraction to a client machine. 
When an unmodified application accesses remote memory through different state-of-the-art remote-memory solutions (\eg, via VFS or VMM), the {\host} transparently handles all aspects of RDMA communication and erasure coding. 
Each client has its own {\host} that handles slab placement through {\copysets}, maintains remote slab-address mapping, performs erasure-coded RDMA read/write. 
{\host} communicates to {\daemon}(s) running on remote memory host machines, performs remote data placement, and ensures resilience. 
As a client's {\host} is responsible for the resiliency of its remote data, the {\host}s do not need to coordinate with each other.

\begin{figure}[t]
	\centering
	\includegraphics[width=\columnwidth]{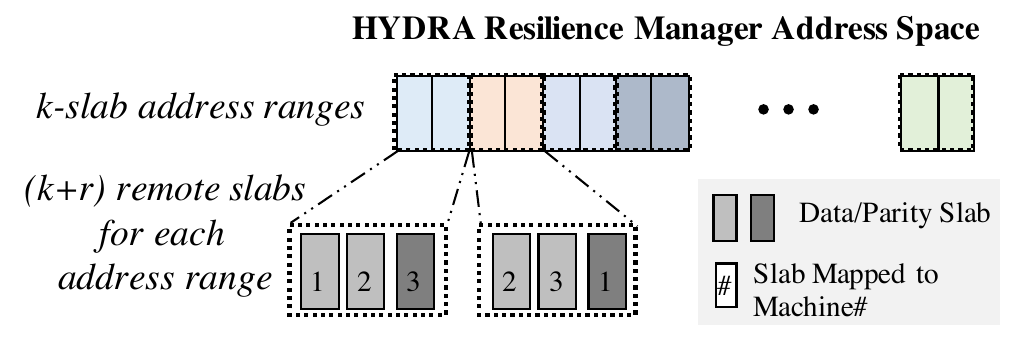}
	\caption{{\name}'s address space is divided into fixed-size address ranges, each spans $(k+r)$ slabs in remote machines; \ie, $k$ for data and $r$ for parity ($k$=2 and $r$=1 in this figure).}
	\label{fig:slab}
\end{figure}

Following the typical $(k, r)$ erasure coding construction, the {\host} divides its remote address space into fixed-size \emph{address ranges}. 
Each address range resides in $(k+r)$ remote \emph{slabs}: $k$ slabs for page data and $r$ slabs for parity (Figure~\ref{fig:slab}). 
Each of the $(k+r)$ slabs of an address range are distributed across $(k+r)$ independent failure domains using \copysets  (\S\ref{sec:analysis}).
Page accesses are directed to the designated $(k+r)$ machines according to the address--slab mapping.
Although remote I/O happens at the page level, the {\host} coordinates with remote {\daemon}s to manage coarse-grained memory slabs to reduce metadata overhead and connection management complexity.

\subsection{{\daemon}}
{\daemon} manages a machine's local memory and exposes them to the remote {\host} in terms of fixed-size ($SlabSize$) memory slabs. 
Different slabs can belong to different machines' {\host}.
During each control period ($ControlPeriod$), the {\daemon} tracks the available memory in its local machine and proactively allocates (reclaims) slabs to (from) remote mapping when memory usage is low (high). 
It also performs slab regeneration during remote failures or corruptions.


\paragraph{Fragmentation in Remote Memory} 
\rev{During the registration of {\daemon}(s), {\host} registers the RDMA memory regions and allocates slabs on the remote machines based on its memory demand. 
Memory regions are usually large (by default, 1GB) and the whole address space is homogeneously splitted. 
Moreover, RDMA drivers guarantee the memory regions are generated in a contiguous physical address space to ensure faster remote-memory access. 
{\name} introduces no additional fragmentation in remote machines.}

\subsection{Failure Model}
\paragraph{Assumptions} 
\rev{In a large remote-memory cluster, (a) remote servers may crash or networks may become partitioned; 
(b) remote servers may experience memory corruption; 
(c) the network may become congested due to background traffic; 
and (d) workloads may have bursty access patterns. 
These can lead to catastrophic application-failures, high tail latencies, or unpredictable performance. 
{\name} addresses all of these uncertainties in its failure domain.}
\rev{Although {\name} withstands a remote-network partition, as there is no local-disk backup, it cannot handle local-network failure. 
In such cases, the application is anyways inaccessible.}

\paragraph{Single vs. Simultaneous Failure} 
\rev{A single node failure means the unavailability of slabs in a remote machine. 
In such an event, all the data or parity allocated on the slab(s) become unavailable. 
As we spread the data and parity splits for a page across multiple remote machines (\S\ref{sec:analysis}), during a single node failure, we assume that only a single data or parity split for that page is being affected.}

\rev{Simultaneous host failures typically occur due to large-scale failures, such as power or network outage that cause multiple machines to become unreachable. 
In such a case, we assume multiple data and/or parity splits for a page become unavailable. 
Note that in both cases, the data is unavailable, but not compromised. 
{\host} can detect the unreachability and communicate to other available {\daemon}(s) on to regenerate specific slab(s).}

%% file: design.tex
\section{Resilient Data Path}
\label{sec:design}

\rev{
{\name} can operate on different resilient modes based on a client's need -- \textbf{(a)} \textit{Failure Recovery}: provides resiliency in the presence of any remote failure or eviction; \textbf{(b)} \textit{Corruption Detection}: only detects the presence of corruption in remote memory; \textbf{(c)} \textit{Corruption Correction}: detects and corrects remote memory corruption; and \textbf{(d)} \textit{EC-only mode}: provides erasure-coded faster remote-memory data path without any resiliency guarantee.
Note that both of the corruption modes by default inherit the \textit{Failure Recovery} mode.}

\rev{
Before initiating the {\host}, one needs to configure Hydra to a specific mode according to the resilience requirements and memory overhead concerns (Table~\ref{tab:guarantee-req}). 
Multiple resilience modes cannot act simultaneously, and the modes do not switch dynamically during runtime.}
%
%
In this section, we present {\name}'s data path design to address the resilience challenges mentioned in \S\ref{subsec:challenge}.

\subsection{{\name} Remote Memory Data Path}
To minimize erasure coding's latency overheads, {\host}'s data path incorporate following design principles.

\begin{figure}[!t]
	\centering
	\subfloat[][Remote Write]{%
		\label{fig:timeline-write}%
		\includegraphics[width=0.48\columnwidth]{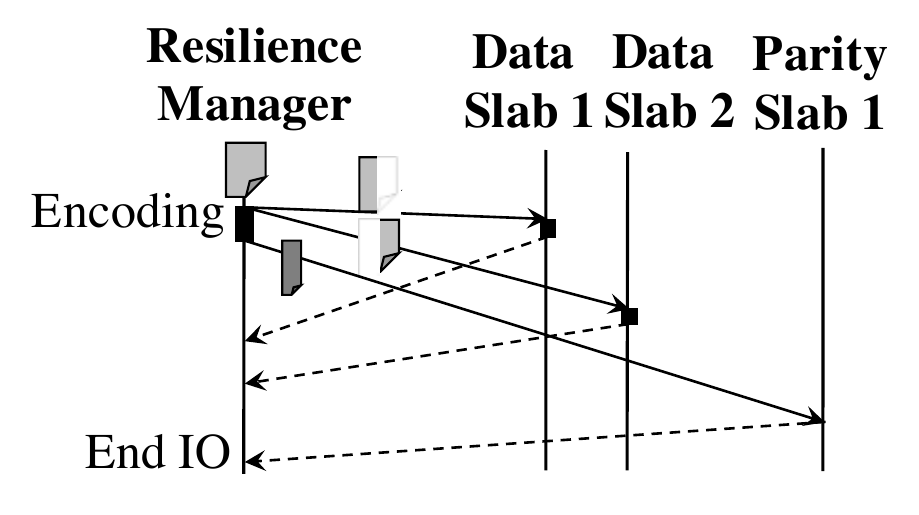}
	}    
	\subfloat[][Remote Read]{%
		\label{fig:timeline-read}%
		\includegraphics[width=0.48\columnwidth]{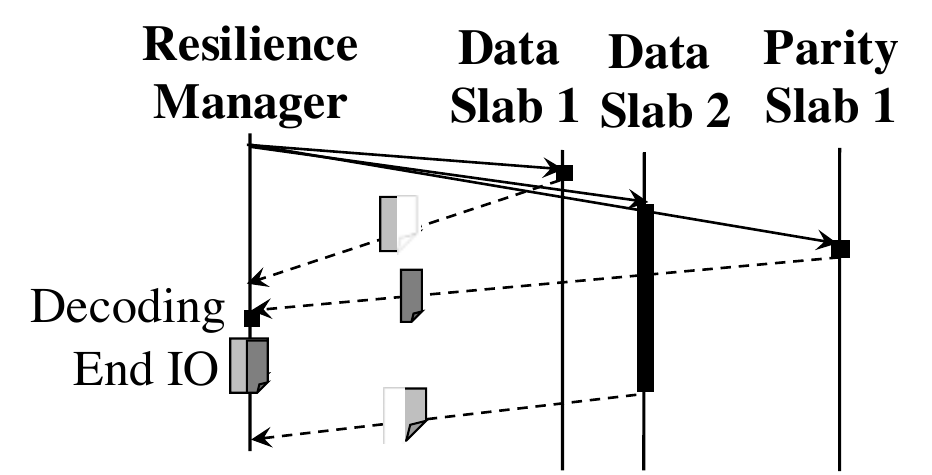}
	}    
	\caption{To handle failures, {\name} (a) first writes data splits, then encodes/writes parity splits to hide encoding latency; 
	(b) reads from $k+\Delta$ slabs to avoid stragglers, finishes with first arrival of $k$ splits.}
	\label{fig:timeline}
	\vspace{1mm}
\end{figure}

\subsubsection{\textbf{Asynchronously Encoded Write}}
\label{subsec:write}
\rev{
To hide the erasure coding latency, existing systems usually perform batch coding where multiple pages are encoded together. 
The encoder waits until a certain number of pages are available.
This idle waiting time can be insignificant compared to disk or slow network (\eg, TCP) access.
However, to maintain the tail latency of a remote I/O  within the single-digit $\mu s$ range, this ``batch waiting'' time needs to be avoided.}

During a remote write, {\host} applies erasure coding within each individual page by dividing it into $k$ splits (for a 4 KB page, each split size is $\frac{4}{k}$KB), encodes these splits using Reed-Solomon (RS) 
codes \cite{reed-solomon} to generate $r$ parity splits. 
Then, it writes these $(k+r)$ splits to different $(k+r)$ slabs that have already been mapped to unique remote machines. 
Each {\host} can have their own choice of $k$ and $r$.
This individual page-based coding decreases latency by avoiding the ``batch waiting'' time. 
Moreover, the {\host} does not have to read unnecessary pages within the same batch during remote reads, which reduces bandwidth overhead.
Distributing remote I/O across many remote machines increases I/O parallelism too.

{\host} sends the data splits first, then it encodes and sends the parity splits asynchronously. 
Decoupling the two hides encoding latency and subsequent write latency for the parities without affecting the resilience guarantee.
In the absence of failure, any $k$ successful writes of the $(k+r)$ allow the page to be recovered.
However, to ensure resilience guarantee for $r$ failures, all $(k+r)$ must be written.
In the {\it failure recovery} mode, a write is considered complete after all $(k+r)$ have been written.
In the {\it corruption correction (detection)} mode, to correct (detect) $\Delta$ corruptions, a write waits for $k+2\Delta+1$ $(k+\Delta)$ to be written.
\rev{If the acknowledgement fails to reach the {\host} due to a failure in the remote machine, the write for that split is considered failed. 
{\host} tries to write that specific split(s) after a timeout period to another remote machine.}
Figure~\ref{fig:timeline-write} depicts the timeline of a page write in the {\it failure recovery mode}. 

\subsubsection{\textbf{Late-Binding Resilient Read}}
\label{subsec:read}
During read, any $k$ out of the $k+r$ splits suffice to reconstruct a page. 
However, in {\it failure recovery} mode, to be resilient in the presence of $\Delta$ failures, during a remote read, {\name} {\host} reads from $k+\Delta$ randomly chosen splits in parallel. 
A page can be decoded as soon as any $k$ splits arrive out of $k+\Delta$.
The additional $\Delta$ reads mitigate the impact of stragglers on tail latency as well.
Figure~\ref{fig:timeline-read} provides an example of a read operation in the {\it failure recovery} mode with $k=2$ and $\Delta=1$, where one of the data slabs (Data Slab 2) is a straggler. 
$\Delta=1$ is often enough in practice.

\begin{table}[!t]
  \centering
  \begin{tabular}{|l|l|l|l|}
  \hline
   \rev{Resilience Mode} & \begin{tabular}[c]{@{}l@{}}\# of \\Errors\end{tabular} & \begin{tabular}[c]{@{}l@{}}Minimum \\\# of Splits\end{tabular}  & \begin{tabular}[c]{@{}l@{}}Memory \\Overhead\end{tabular}\\ 

  \hline
  \rev{Failure Recovery}              & $r$   & $k$       & $1+\frac{r}{k}$ \\
  \hline
  \rev{Corruption Detection}      & $\Delta$ & $k+\Delta$   & $1+\frac{\Delta}{k}$       \\
  \hline
  \rev{Corruption Correction}  & $\Delta$    & $k+2\Delta+1$    & $1+\frac{2\Delta+1}{k}$   \\
  \hline
    \rev{EC-only}  & \rev{--}    & \rev{$k$}    & \rev{$1+\frac{r}{k}$}  \\
    \hline
  \end{tabular}
  \caption{Min. number of splits needs to be written to/read from remote machines for resilience during a remote I/O.
  }
  \label{tab:guarantee-req}
  \vspace{-4mm}
\end{table}

\rev{If simply ``detect and discard corrupted memory" is enough for any application, one can configure {\name} with {\it corruption detection} mode and avoid the extra memory overhead of {\it corruption correction} mode.
In {\it corruption detection} mode, before decoding a page, the {\host} waits for $k+\Delta$ splits to arrive to \emph{detect} $\Delta$ corruptions.
After the detection of a certain amount of corruptions, {\host} marks the machine(s) with corrupted splits as probable erroneous machines, initiates a background slab recovery operation, and avoids them during future remote I/O.}

\rev{To correct the error, in {\it corruption correction} mode, when an error is detected}, it requests additional $\Delta+1$ reads from the rest of the $k+r$ machines. 
Otherwise, the read completes just after the arrival of the $k+\Delta$ splits. 
If the error rate for a remote machine exceeds a user-defined threshold ($ErrorCorrectionLimit$), subsequent read requests involved with that machine initiates with $k+2\Delta+1$ split requests as there is a high probability to reach an erroneous machine.
This will reduce the wait time for additional $\Delta+1$ reads. 
This continues until the error rate for the involved machine gets lower than the $ErrorCorrectionLimit$. 
If this continues for long and/or the error rate goes beyond another threshold ($SlabRegenerationLimit$), {\host} initiates the slab regeneration request. 

\rev{One can configure {\name} with {\it EC-only} mode to access erasure-coded remote memory and benefit from the fast data path without any resiliency guarantee. 
In this mode, a remote I/O completes just after writing/reading any k splits.}
Table~\ref{tab:guarantee-req} summarizes the minimum number of splits requires to write/read during a remote I/O operation to provide resiliency in different modes.

\paragraph{\textbf{Overhead of Replication}}
To remain operational after $r$ failures, in-memory replication requires at least $r+1$ copies of an entire 4 KB page, and hence the memory overhead is $(r+1)\times$. 
However, a remote I/O operation can complete just after the confirmation from one of the $r+1$ machines.  
To detect and fix $\Delta$ corruptions, replication needs $\Delta+1$ and $2\Delta+1$  copies of the \emph{entire} page, respectively. 
Thus, to provide the correctness guarantee over $\Delta$  corruptions, replication needs to wait until it writes to or reads from at least $2\Delta+1$ of the replicas along with a memory overhead of $(2\Delta+1)\times$.


\subsubsection{\textbf{Run-to-Completion}}
As {\host} divides a 4 KB page into $k$ smaller pieces, RDMA messages become smaller. 
In fact, their network latency decrease to the point that run-to-completion becomes more beneficial than a context switch. 
Hence, to avoid interruption-related overheads, the remote I/O request thread waits until the RDMA operations are done.

\subsubsection{\textbf{In-Place Coding}}
\begin{figure}[t]
	\centering
	\subfloat[][Remote Write]{
	\includegraphics[scale=0.34]{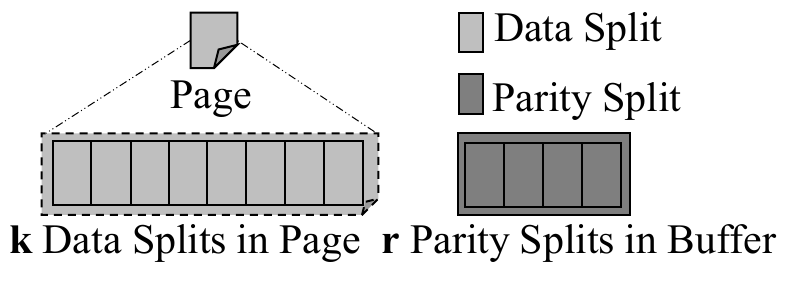}
	\label{fig:inplace-coding-write}
	}
	\subfloat[][Remote Read]{
	\includegraphics[scale=0.34]{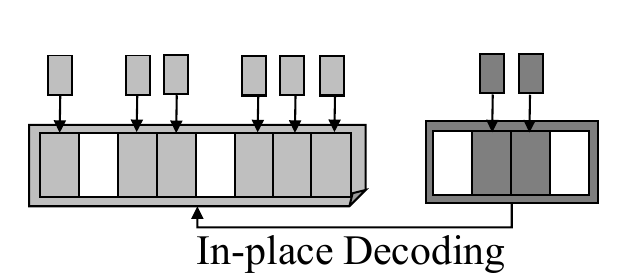}
	\label{fig:inplace-coding-read}
	}
	\caption{{\name} performs in-place coding with an extra buffer of $r$ splits to reduce the data-copy latency.}
	\label{fig:inplace-coding}
\end{figure}

To reduce the number of data copies, {\name} {\host} uses in-place coding with an extra buffer of $r$ splits.
During a write, the data splits are always kept in-page while the encoded $r$ parities are put into the buffer (Figure~\ref{fig:inplace-coding-write}).
Likewise, during a read, the data splits arrive at the page address, and the parity splits find their way into the buffer (Figure~\ref{fig:inplace-coding-read}). 

In the failure recovery mode, a read can complete as soon as any $k$ valid splits arrive.
Corrupted/straggler data split(s) can arrive late and overwrite valid page data.
To address this, as soon as {\name} detects the arrival of $k$ valid splits, it deregisters relevant RDMA memory regions.
It then performs decoding and directly places the decoded data in the page destination. 
Because the memory region has already been deregistered, any late data split cannot access the page.
During all remote I/O, requests are forwarded directly to RDMA dispatch queues without additional copying.  

\subsection{Handling Uncertainties}

\paragraph{\textbf{Remote Failure}}
{\name} uses reliable connections (RC) for all RDMA communication.
Hence, we consider unreachability due to machine failures/reboots or network partition as the primary cause of failure. 
When a remote machine becomes unreachable, the {\host} is notified by the RDMA connection manager. 
Upon disconnection, it processes all the in-flight requests in order first. 
For ongoing I/O operations, it resends the I/O request to other available machines.
Since RDMA guarantees strict ordering, in the read-after-write case, read requests will arrive at the same RDMA dispatch queue after write requests; hence, read requests will not be served with stale data. 
Finally, {\name} marks the failed slabs and future requests are directed to the available ones.
\rev{If the {\daemon} in the failed machine revives and communicates later, {\name} reconsiders the machine for further remote I/O.}


%

\paragraph{\textbf{Adaptive Slab Allocation/Eviction}}
{\daemon} allocates memory slabs for {\host}s as well as proactively frees/evicts them to avoid local performance impacts (Figure~\ref{fig:control}). 
It periodically monitors local memory usage and maintains a headroom to provide enough memory for local applications.
When the amount of free memory shrinks below the headroom (Figure~\ref{fig:control-high}), the {\daemon} first proactively frees/evicts slabs to ensure local applications are unaffected. 
%
\rev{
To find the eviction candidates, we avoid random selection as it has a higher likelihood of evicting a busy slab. 
Rather, we uses the decentralized batch eviction algorithm \cite{infiniswap} to select the least active slabs. 
To evict $E$ slabs, we contact $(E+E')$ slabs (where $E' \le E$) and find the least-frequently-accessed slabs among them. 
This doesn't require to maintain a global knowledge or search across all the slabs.} 

When the amount of free memory grows above the headroom (Figure~\ref{fig:control-low}), the {\daemon} first attempts to make the local {\host} to reclaim its pages from remote memory and unmap corresponding remote slabs. 
Furthermore, it proactively allocates new, unmapped slabs that can be readily mapped and used by remote {\host}s. 

\paragraph{\textbf{Background Slab Regeneration}}
The {\daemon} also regenerates unavailable slabs -- marked by the {\host} -- in the background. 
During regeneration, writes to the slab are disabled to prevent overwriting new pages with stale ones; reads can still be served without interruption.

{\name} {\host} uses the placement algorithm to find a new regeneration slab in a remote {\daemon} with a lower memory usage. 
It then hands over the task of slab regeneration to that {\daemon}. 
The selected {\daemon} decodes the unavailable slab by directly reading the $k$ randomly-selected remaining valid slab for that address region. 
Once regeneration completes, it contacts the {\host} to mark the slab as available. 
Requests thereafter go to the regenerated slab. 


%% file: analysis.tex
\section{\copysets for High Availability}
\label{sec:analysis}

{\name} uses \copysets, a novel coding group placement scheme to perform load-balancing while reducing the probability of data loss. 
%
%
Prior works show orders-of-magnitude more frequent data loss due to events 
causing multiple nodes to fail simultaneously than data loss due to independent 
node failures~\cite{unavailability-study,tiered}.
Several scenarios can cause multiple servers to fail or become unavailable simultaneously, such as 
network partitions, partial power outages, and software bugs.
For example, a power outage can cause 0.5\%-1\% machines 
to fail or go offline concurrently \cite{copyset}.
In case of \name, data loss will happen if a concurrent failure kills more than $r+1$ 
of $(k+r)$ machines for a particular \emph{coding group}.

We are inspired by copysets, a scheme for preventing data loss under correlated failures in replication~\cite{copyset,tiered}, which constraints the number of replication groups, in order to \emph{reduce the frequency of data loss events}.
Using the same terminology as prior work, we define each unique set of $(k+r)$ servers within a coding group as a $copyset$.
The number of copysets in a single coding group will be: ${k+r}\choose {r+1}$.
For example, in an (8+2) configuration, where nodes are numbered $1,2,\ldots,10$, 
the 3 nodes that will cause failure if they fail at the same time (\ie, copysets) 
will be every 3 combinations of 10 nodes:
$(1, 2, 3), (1, 2, 4), \ldots, (8, 9, 10)$, and the total number of copysets will be $\binom{10}{3} = 120$.

\begin{figure}[t]
	\centering
	\includegraphics[width=\columnwidth]{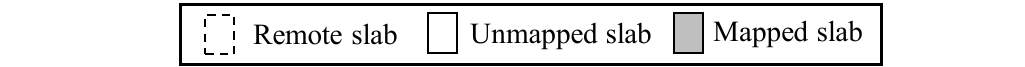}\\[\dimexpr -0.3cm]
	\subfloat[][High memory pressure]{%
		\label{fig:control-high}%
		\includegraphics[width=1.6in]{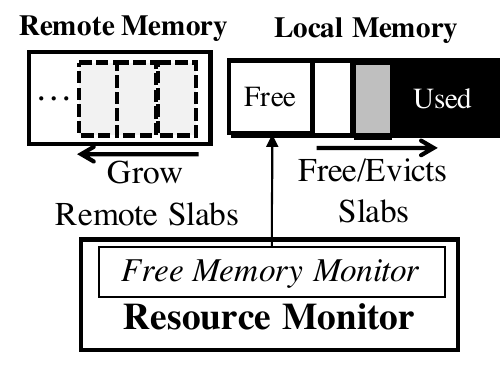}
	}
	\subfloat[][Low memory pressure]{%
		\label{fig:control-low}%
		\includegraphics[width=1.6in]{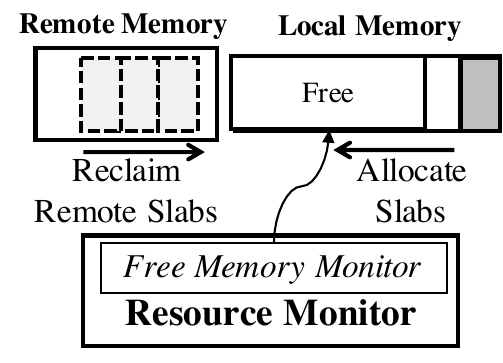}
	}
	\caption{{\daemon} proactively allocates memory for remote machines and frees local memory pressure.}
	\label{fig:control}
\end{figure}

For a data loss event impacting exactly $r+1$ random nodes simultaneously, 
the probability of losing data of a single specific coding group: $\Prob{Group}$ = $\frac{\text{Num. of Copysets in Coding Group}}{\text{Total Copysets}}$ =  $\frac{\binom {k+r}{r+1}}{ \binom{N}{r+1}}$, where $N$ is the total number of servers.

In a cluster with more than $(k+r)$ servers, we need to use more than one coding group.
However, if each server is a member of a single coding group, hot spots can occur if one or more members of that group are overloaded.
Therefore, for load-balancing purposes, a simple solution is to allow each server to be a member of multiple coding groups, in case
some members of a particular coding group are over-loaded at the time of online coding. 

Assuming we have $G$ \emph{disjoint} coding groups,
and the correlated failure rate is $f$\%, 
the total probability of data loss is: 
$1-(1-\Prob{Group} \cdot G)^{\binom{N \cdot f}{r+1}}$. 
We define disjoint coding groups where the groups do not share any copysets; or in other words, they do not overlap by more than $r$ nodes.

\paragraph{\textbf{Strawman: Multiple Coding Groups per Server}}

In order to equalize load, we consider a scheme where each slab forms a coding group with the least-loaded nodes in the cluster at coding time.
We assume the nodes that are least loaded at a given time
are distributed randomly, and the number of slabs per server is $S$.
When $S\cdot(r+k)\ll N$, the coding groups are highly likely to 
be disjoint~\cite{copyset}, and the number of groups is equal to:
$G=\dfrac{N \cdot S}{k+r}$.

We call this placement strategy the {\em EC-Cache scheme}, as it produces a random coding group placement used by the prior state-of-the-art in-memory erasure coding system, EC-Cache~\cite{eccache}.
In this scheme, with even a modest number of slabs per server,
a high number of combinations of $r+1$ machines will be a
copyset. In other words, even a small
number of simultaneous node failures in the cluster will result in data loss. 
When the number of slabs per server is high, almost every combination of only 
$r+1$ failures across the cluster will cause data loss.
Therefore, to reduce the probability of data loss, we need to minimize the number of copysets, while achieving sufficient load balancing.

\paragraph{\textbf{\copysets: Reducing Copysets for Erasure Coding}}

To this end, we propose \copysets, a novel load-balancing scheme, which reduces the number of copysets for distributed erasure coding.
Instead of having each node participate in several coding groups like in EC-Cache, in our scheme, each server belongs to a single, \emph{extended} coding group.
At time of coding, $(k+r)$ slabs will still be considered together,
but the nodes participating in the coding group are chosen from a set of $(k+r+l)$ 
nodes, where $l$ is the load-balancing factor. The nodes chosen within the extended 
group are the least loaded ones.
While extending the coding group increases the number of copysets (instead of ${k+r}\choose {r+1}$ copysets, now each
extended coding group creates ${k+r+l}\choose {r+1}$ copysets, while the number of groups is $G=\dfrac{N}{k+r+l}$), it still has a significantly lower probability
of data loss than having each node belong to multiple coding groups.
\name uses \copysets as its load balancing and slab placement policy. We evaluate it in Section~\ref{sec:eval-availability}.

\paragraph{{\bf Tradeoff}} Note that while \copysets reduces the probability of data loss, it does not reduce the expected
amount of data lost over time. In other words, it reduces the number of data loss events,
but each one of these events will have a proportionally higher magnitude of data loss (\ie, more slabs
will be affected)~\cite{copyset}.
Given that our goal with \name is high availability,
we believe this is a favorable trade off. For example, providers often provide an
availability SLA, that is measured by the service available time (\eg, the service is available 99.9999\% of the time).
\copysets would optimize for such an SLA, by minimizing the frequency of unavailability events.

%% file: implementation.tex
\section{Implementation}
\label{sec:implementation}

{\host} is implemented as a loadable kernel module for Linux kernel 4.11 or later. 
Kernel-level implementation facilitates its deployment as an underlying block device for different remote-memory systems \cite{remote-regions, infiniswap, legoos}. 
We integrated {\name} with two remote-memory systems: {\is}, a disaggregated VMM and Remote Regions, a disaggregated VFS.
All I/O operations (\eg, slab mapping, memory registration, RDMA posting/polling, erasure coding) are independent across threads and processed without synchronization.
All RDMA operations use RC and one-sided RDMA verbs (RDMA WRITE/READ). 
Each {\host} maintains one connection for each active remote machine. 
For erasure coding, we use x86 AVX instructions and the ISA library \cite{isa} that achieves over 4 GB/s encoding throughput per core for (8+2) configuration in our evaluation platform. 

{\daemon} is implemented as a user-space program. 
It uses RDMA SEND/RECV for all control messages.

%% file: evaluation.tex
\section{Evaluation}
\label{sec:eval}
We evaluate {\name} on a 50-machine 56 Gbps InfiniBand CloudLab cluster against {\is} \cite{infiniswap}, Leap~\cite{leap} (disaggregated VMM) and Remote Regions \cite{remote-regions} (disaggregated VFS). 
Our evaluation addresses the following questions:
\begin{denseitemize}
	\item Does it improve the resilience of cluster memory? (\S\ref{sec:eval-resilience})
	
	\item Does it improve the availability? (\S\ref{sec:eval-availability})

	\item 
	What is its overhead and sensitivity to parameters? (\S\ref{sec:eval-sensitivity})

	\item How much TCO savings can we expect? (\S\ref{sec:eval-tco})
	
	\item What is its benefit over a persistent memory setup? (\S\ref{sec:pm-backup})
\end{denseitemize}

\paragraph{\textbf{Methodology}}
Unless otherwise specified, we use $k$=8, $r$=2, and $\Delta$=1, targeting $1.25\times$ memory and bandwidth overhead.
We select $r$=2 because late binding is still possible even when one of the remote slab fails.
The additional read $\Delta$=1 incurs $1.125\times$ bandwidth overhead during reads.
We use 1GB $SlabSize$, 
The additional number of choices for eviction $E'=2$.
Free memory headroom is set to 25\%, and the control period is set to 1 second. 
Each machine has 64 GB of DRAM and 2$\times$ Intel Xeon E5-2650v2 with 32 virtual cores.

We compare {\name} against the following alternatives:
\begin{denseitemize}
	\item \textbf{SSD Backup}: Each page is backed up in a local SSD for the minimum $1\times$ remote memory overhead. 
		We consider both disaggregated VMM and VFS systems.
	\item \textbf{Replication}: We directly write each page over RDMA to two 
remote machines' memory for a $2\times$ overhead. 

	\item \textbf{EC-Cache w/ RDMA}: Implementation of the erasure coding scheme in EC-Cache \cite{eccache}, but implemented on RDMA.
\end{denseitemize}

\paragraph{\textbf{Workload Characterization}} 
Our evaluation consists of both micro-benchmarks and cluster-scale evaluations with real-world applications and workload combinations.
\begin{denseitemize}
  \item We use TPC-C~\cite{tpc-c} on VoltDB \cite{voltdb}. 
  We perform 5 different types of transactions to simulate an order-entry environment. 
  We set 256 warehouses and 8 sites and run 2 million transactions. Here, the peak memory usage is 11.5 GB.
  \item We use Facebook's ETC, SYS workloads \cite{fb-memcached-workload} on Memcached \cite{memcached}.
  First, we use 10 million SETs to populate the Memcached server. 
  Then we perform another 10 million operations (for ETC: 5\% SETs, 95\% GETs, for SYS: 25\% SETs, 75\% GETs). 
  The key size is 16 bytes and 90\% of the values are evenly distributed between 16--512 bytes. 
   Peak memory usages are 9 GB for ETC  and 15 GB for SYS.
     \item We use PageRank on PowerGraph~\cite{powergraph} and Apache Spark / GraphX~\cite{graphx} to measure the influence of Twitter users on followers on a graph with 11 million vertices~\cite{twitter-graph-workload}. 
     Peak memory usages are 9.5 GB and 14 GB, respectively.
\end{denseitemize}

\begin{figure}[!t]
	\centering
	\subfloat[][Disaggregated VMM Latency]{%
		\label{fig:lat-vmm}%
		\includegraphics[scale=0.37]{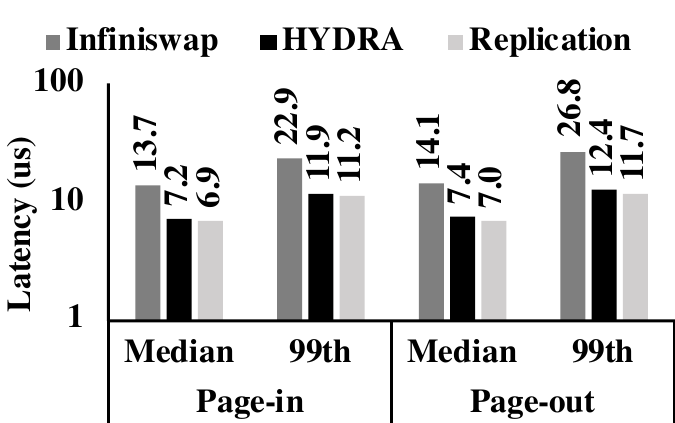}
	}
	\subfloat[][Disaggregated VFS Latency]{%
		\label{fig:lat-vfs}%
		\includegraphics[scale=0.37]{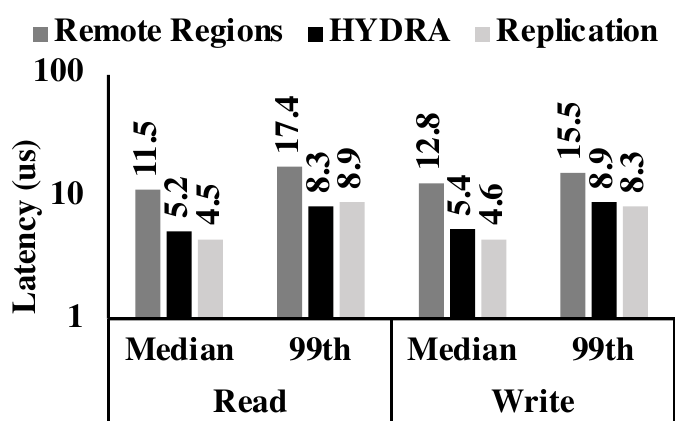}
	}
	\caption{{\name} provides better latency characteristics during both disaggregated VMM and VFS operations.
	}
	\label{fig:lat}
\end{figure}

\begin{figure}[t]
	\centering
	\subfloat[][Remote Read]{%
		\label{fig:dataplane-breakdown-read}%
		\includegraphics[width=0.5\columnwidth]{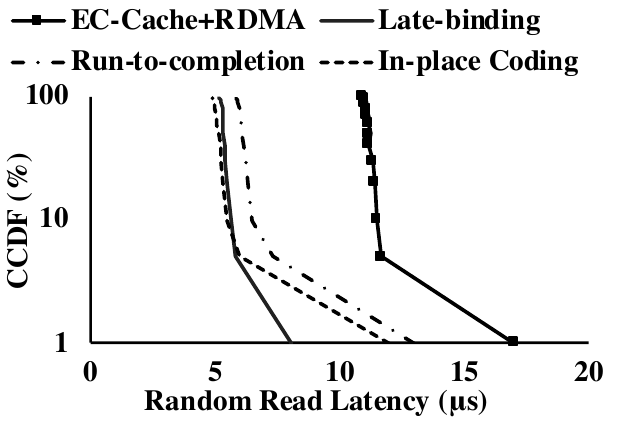}
	}    
	\subfloat[][Remote Write]{%
		\label{fig:dataplane-breakdown-write}%
		\includegraphics[width=0.5\columnwidth]{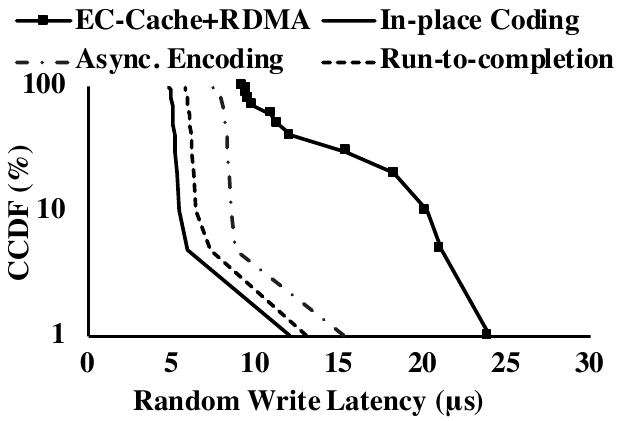}
	}    
	\caption{{\name} latency breakdown through CCDF.
	}
	\label{fig:dataplane-breakdown}
\end{figure}

\subsection{Resilience Evaluation}
\label{sec:eval-resilience}
We evaluate {\name} both in the presence and absence of failures with microbenchmarks and real-world applications.

\subsubsection{\textbf{Latency Characteristics}}
First, we measure {\name}'s latency characteristics with micro-benchmarks in the absence of failures.
Then we analyze the impact of its design components.

\paragraph{\textbf{Disaggregated VMM Latency}}
We use a simple application with its working set size set to 2GB. 
It is provided 1GB memory to ensure that 50\% of its memory accesses cause paging.
While using disaggregated memory for remote page-in, {\name} improves page-in latency over {\is} with SSD backup by 1.79$\times$ at median and 1.93$\times$ at the 99th percentile. 
Page-out latency is improved by 1.9$\times$ and 2.2$\times$ over {\is} at median and 99th percentile, respectively. 
Replication provides at most 1.1$\times$ improved latency over {\name}, while incurring $2\times$ memory and bandwidth overhead (Figure~\ref{fig:lat-vmm}).

\paragraph{\textbf{Disaggregated VFS Latency}}
We use \texttt{fio} \cite{fio} to generate one million random read/write requests of 4 KB block I/O.
During reads, {\name} provides improved latency over Remote Regions by 2.13$\times$ at median and 2.04$\times$ at the 99th percentile. 
During writes, {\name} also improves the latency over Remote Regions by 2.22$\times$ at median and 1.74$\times$ at the 99th percentile.
Replication has a minor latency gain over {\name}, improving latency by at most 1.18$\times$ (Figure~\ref{fig:lat-vfs}).


\paragraph{\textbf{Benefit of Data Path Components}}
Erasure coding over RDMA (\ie, EC-Cache with RDMA) performs worse than disk backup due to its coding overhead. 
Figure~\ref{fig:dataplane-breakdown} shows the benefit of  {\name}'s data path components to reduce the latency. 

\begin{denseenum}	
	\item Run-to-completion avoids interruptions during remote I/O, reducing the median read and write latency by 51\%.
	
	\item In-place coding saves additional time for data copying, which substantially adds up in remote-memory systems, reducing 28\% of the read and write latency.
	
	\item Late binding specifically improves the tail latency during remote read by 61\% by avoiding stragglers. 
	The additional read request increases the median latency only by 6\%. 
	
		\item Asynchronous encoding hides erasure coding overhead during writes, reducing the median write latency by 38\%.
\end{denseenum}

 \begin{figure}[t]
 	\subfloat[][Read breakdown]{
 		\label{fig:breakdown-read}%
 		\includegraphics[width=0.5\columnwidth]{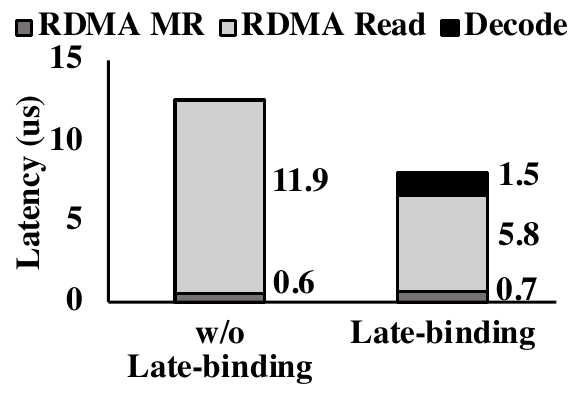}
 	}	
 	\subfloat[][Write breakdown]{
 		\label{fig:breakdown-write}%
 		\includegraphics[width=0.5\columnwidth]{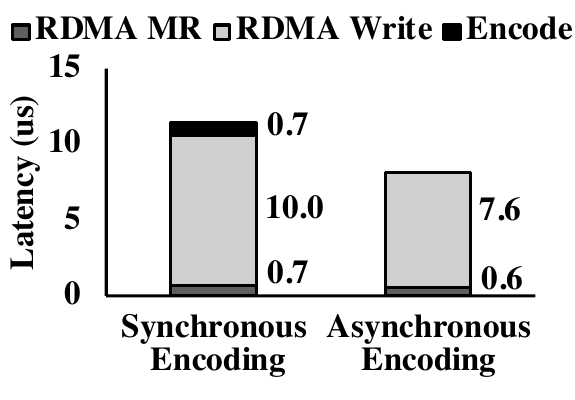}
 	}
 	\caption{{\name} latency breakdown at the 99$^{th}$ percentile.}
 	\label{fig:breakdown}
 \end{figure}

\begin{figure}[!t]
	\subfloat[][Background network flow]{
		\label{fig:lat-contention}%
		\includegraphics[width=0.5\columnwidth]{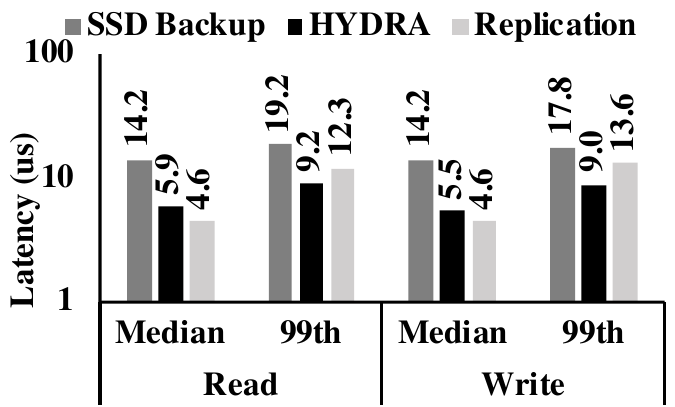}
	}
	\subfloat[][Remote Failures]{
		\label{fig:lat-failure}%
		\includegraphics[width=0.5\columnwidth]{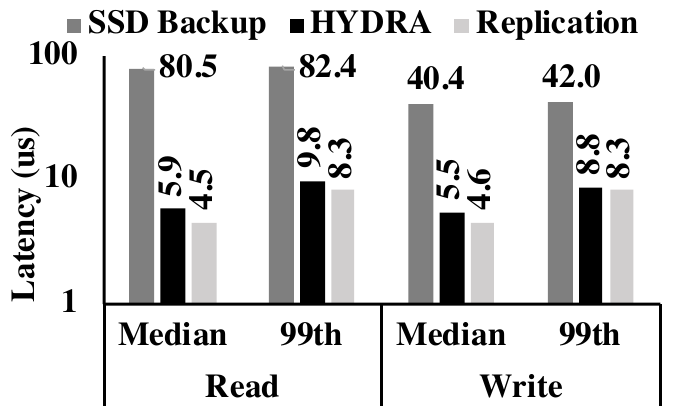}
	}	
	\caption{Latency in the presence of uncertainty events.
	}
	\label{fig:lat-uncertain}
	\vspace{2mm}
\end{figure}

\paragraph{{\bf Tail Latency Breakdown}}
 The latency of {\name} consists of the time for 
 (i) RDMA Memory Registration (MR), 
 (ii) actual RDMA read/write, and 
 (iii) erasure coding.
Even though decoding a page takes about $1.5\mu s$, late binding effectively improves the tail latency by 1.55$\times$ (Figure~\ref{fig:breakdown-read}). 
 During writes, asynchronous encoding hides encoding latency and latency impacts of straggling splits, improving tail latency by 1.34$\times$ w.r.t. synchronous encoding (Figure~\ref{fig:breakdown-write}).
 At the presence of corruption ($r=3$), accessing extra splits increases the tail latency by 1.51$\times$ and 1.09$\times$ for reads and writes, respectively.

\subsubsection{\textbf{Latency Under Failures}}

\paragraph{\textbf{Background Flows}}
We generate RDMA flows on the remote machine constantly sending 1 GB messages.
Unlike SSD backup and replication, {\name} ensures consistent latency due to late binding (Figure~\ref{fig:lat-contention}).
{\name}'s latency improvement over SSD backup is 1.97--2.56$\times$. It even outperforms replication at the  tail read (write) latency by 1.33$\times$ (1.50$\times$).

\paragraph{\textbf{Remote Failures}}
Both read and write latency are disk-bound when it's necessary to access the backup SSD (Figure~\ref{fig:lat-failure}). 
{\name} reduces latency over SSD backup by 8.37--13.6$\times$ and 4.79--7.30$\times$ during remote read and write, respectively.
Furthermore, it matches the performance of replication.

\begin{table}[!t]
	\footnotesize
	\centering
	\begin{tabular}{|c|r|r|r||r|r||r|r|}            
		\hline	
		\multicolumn{2}{|c|}{\multirow{2}{*}{}} &
		\multicolumn{2}{c||}{\begin{tabular}[c]{@{}c@{}}TPS/OPS\\(thousands)\end{tabular}} &
		\multicolumn{4}{c|} {Latency (ms)}\\
		\multicolumn{2}{|c|}{\multirow{2}{*}{}} &
		\multicolumn{2}{c||}{\multirow{2}{*}{}} &
		\multicolumn{2}{c||}{ 50th} & \multicolumn{2}{c|}{ 99th}\\		
		\multicolumn{2}{|c|}{}& HYD & REP & HYD & REP & HYD & REP \\	
		\hline
		\multirow{3}{*}{VoltDB}
		& 100\% & 39.4  & 39.4 & 52.8 & 52.8 & 134.0 & 134.0\\
		& 75\%	& 36.1  & 35.3 & 56.3 & 56.1 & 142.0 & 143.0\\
		& 50\%  & 32.3  & 34.0 & 57.8 & 59.0 & 161.0 & 168.0\\
		\hline
		\multirow{3}{*}{ETC}	 
		& 100\% & 123.0 & 123.0 & 123.0 & 123.0 & 257.0 & 257.0 \\
		& 75\%  & 119.0 & 125.0 & 120.0 & 121.0 & 255.0 & 257.0\\
		& 50\%  & 119.0 & 119.0 & 118.0 & 122.0 & 254.0 & 264.0 \\
		\hline
		\multirow{3}{*}{SYS}
		& 100\% & 108.0 & 108.0 & 125.0 & 125.0 & 267.0 & 267.0\\
		& 75\%  & 100.0 & 104.0 & 120.0 & 125.0 & 262.0 & 305.0\\
		& 50\%  & 101.0 & 102.0 & 117.0 & 123.0 & 257.5 & 430.0\\	 
		\hline
	\end{tabular}
	\caption{{\name} (HYD) provides similar performance to replication (REP) for VoltDB and Memcached (ETC and SYS). 
		Higher is better for  throughput; Lower is better for latency.    
	}
	\label{tab:apps-ec}
	\vspace{-4mm}
\end{table}

\subsubsection{\textbf{Application-Level Performance}}
\label{subsec:apps}
We now focus on {\name}'s benefits for real-world memory-intensive applications and compare it with that of SSD backup and replication. 
We consider container-based application deployment \cite{borg} and run each application in an \texttt{lxc} container with a memory limit to fit 100\%, 75\%, 50\% of the peak memory usage for each application. 
For 100\%, applications run completely in memory. 
For 75\% and 50\%, applications hit their memory limits and performs remote I/O via {\name}. 

We present {\name}'s application-level performance against replication (Table~\ref{tab:apps-ec} and Table~\ref{tab:apps-ec-graph}) to show that it can achieve similar performance with a lower memory overhead even in the absence of any failures. 
For brevity, we omit the results for SSD backup, which performs much worse than both {\name} and replication -- albeit with no memory overhead. 

For VoltDB, when half of its data is in remote memory, {\name} achieves 0.82$\times$ throughput and almost transparent latency characteristics compared to the fully in-memory case. 

\begin{table}[!t]
\vspace{2mm}
\footnotesize
\centering
	\begin{tabular}{|c|l|r|r||r|r|r|}
	\hline
 & \multicolumn{3}{c||}{\begin{tabular}[c]{@{}c@{}}Apache Spark/GraphX\\ Completion Time (s)\end{tabular}} & \multicolumn{3}{c|}{\begin{tabular}[c]{@{}c@{}}PowerGraph\\ Completion Time (s)\end{tabular}} \\ \cline{2-7}
 & 100\% & 75\% & 50\% & 100\% & 75\% & 50\% \\ 
\hline
Hydra & 77.91 & 105.41 & 191.93 & 73.10 & 66.90 & 68.00 \\ 
\hline
Replication & 77.91 & 91.89 & 195.54 & 73.10 & 73.30 & 73.70 \\ 
\hline
\end{tabular}
\caption{{\name} also provides similar completion time to replication for graph analytic applications.}
\label{tab:apps-ec-graph}
\vspace{-4mm}
\end{table}

\begin{figure*}[t]
	\centering    
	\subfloat[][Remote failure]{%
		\label{fig:failure}
		\begin{minipage}{0.24\textwidth}
			\includegraphics[width=\textwidth]{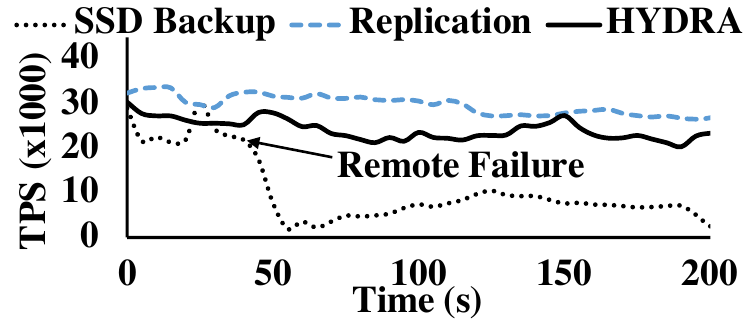}
		\end{minipage}
	}
	\subfloat[][Remote Network Load]{%
		\label{fig:load}%
		\begin{minipage}{0.24\textwidth}
			\includegraphics[width=\textwidth]{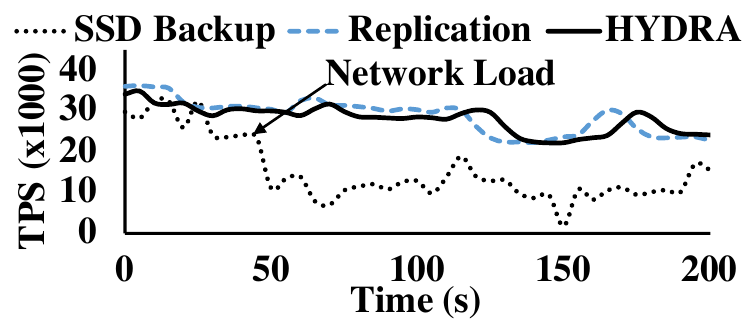}
		\end{minipage}
	}
	\subfloat[][Request Burst]{
		\label{fig:burst}
		\begin{minipage}{0.24\textwidth}
			\includegraphics[width=\textwidth]{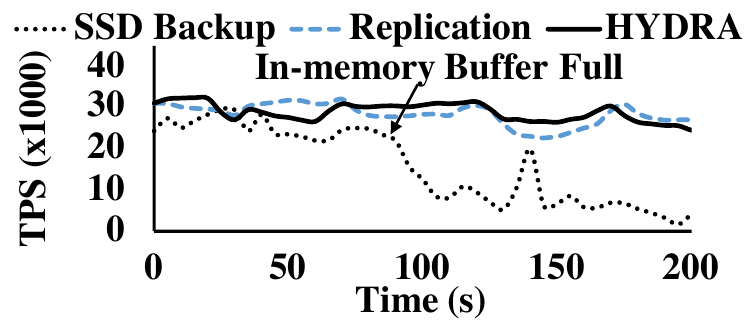}
		\end{minipage}
	}
	 \subfloat[][Page Corruption]{%
	 	\label{fig:corrupt}%
	 	\begin{minipage}{0.24\textwidth}
	 		\includegraphics[width=\textwidth]{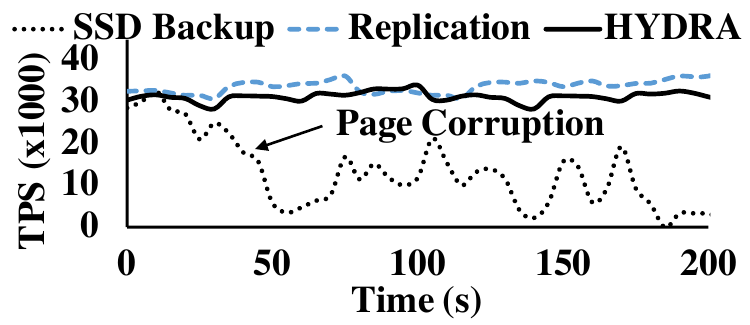}
	 	\end{minipage}
	 }    
	\caption{{\name} throughput with the same setup in Figure~\ref{fig:motivation}.
	}
	\label{fig:result-ec}
\end{figure*}

For Memcached, at 50\% case, {\name} achieves 0.97$\times$ throughput with read-dominant ETC workloads and 0.93$\times$ throughput with write-intensive SYS workloads compared to the 100\% scenario.
Here, latency overhead is almost zero.

For graph analytics, {\name} could achieve almost transparent application performance for PowerGraph; thanks to its optimized heap management. 
However, it suffers from increased job completion time for GraphX due to massive thrashing of in-memory and remote memory data -- the 14 GB working set oscillates between paging-in and paging-out. 
This causes bursts of RDMA reads and writes. 
Even then, Hydra outperforms {\is} with SSD backup by $8.1\times$.
Replication does not have significant gains over {\name}.

\paragraph{Performance with Leap}
\rev{
{\name}'s drop-in resilience mechanism is orthogonal to the functionalities of remote-memory frameworks. 
To observe {\name}'s benefit even with faster in-kernel lightweight remote-memory data path, we integrate it to Leap~\cite{leap} and run VoltDB and PowerGraph with 50\% remote-memory configurations.}

\rev{Leap waits for an interrupt during a 4KB remote I/O, whereas {\name} splits a 4KB page into smaller chunks and performs asynchronous remote I/O. 
Note that RDMA read for 4KB-vs-512B is 4$\mu$s-vs-1.5$\mu$s.
With self-coding and run-to-completion, {\name} provides competitive performance guarantees as Leap for both VoltDB (0.99$\times$ throughput) and PowerGraph (1.02$\times$ completion time) in the absence of failures.}

\begin{figure}[!t]
	\centering
	\includegraphics[width=\columnwidth]{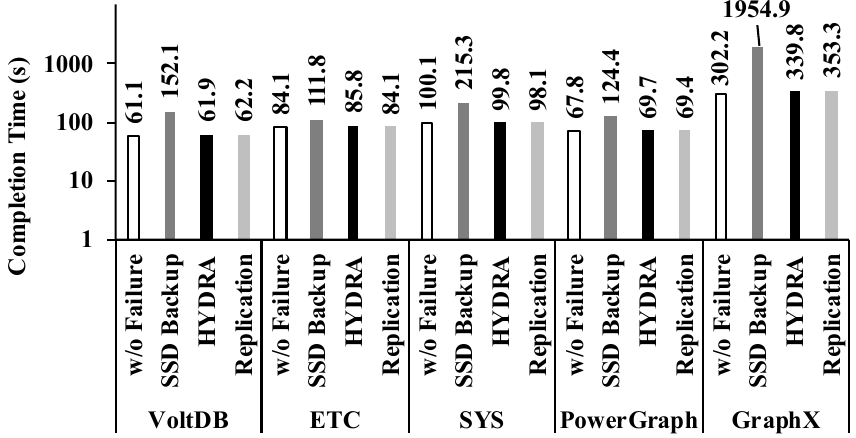}
	\caption{{\name} provides transparent completions in the presence of failure.
		Note that the Y-axis is in log scale.}
	\label{fig:failure-app}
\end{figure}

\subsubsection{\textbf{Application Performance Under Failures}}
\label{subsubsec:app-failure}
Now we analyze {\name}'s performance in the presence of failures and compare against the alternatives. 
In terms of impact on applications, we first go back to the scenarios discussed in Section~\ref{subsec:uncertainties} regarding to VoltDB running with 50\% memory constraint.
Except for the corruption scenario where we set $r$=3, we use {\name}'s default parameters.
At a high level, we observe that {\name} performs similar to replication with 1.6$\times$ lower memory overhead (Figure~\ref{fig:result-ec}).

Next, we start each benchmark application in 50\% settings and introduce one remote failure while it is running.
\rev{We select a {\daemon} with highest slab activities and kill it. 
We measure the application's performance while the {\host} initiates the regeneration of affected slabs. }

{\name}'s application-level performance is transparent to the presence of remote failure. 
Figure~\ref{fig:failure-app} shows {\name} provides almost similar completion times to that of replication at a lower memory overhead in the presence of remote failure. 
In comparison to SSD backup, workloads experience 1.3--5.75$\times$ lower completion times using {\name}.
{\name} provides similar performance at the presence of memory corruption. 
Completion time gets improved by 1.2--4.9$\times$ w.r.t. SSD backup.

\subsection{Availability Evaluation}
\label{sec:eval-availability}
In this section, we evaluate {\name}'s availability and load balancing characteristics in large clusters.

\subsubsection{\textbf{Analysis of \copysets}}

We compare the availability and load balancing of \name with EC-Cache and power-of-two-choices~\cite{power-of-2}.
In \copysets, each server is attached to a disjoint coding group. During encoded write, the $(k+r)$ least loaded nodes are chosen from a subset of the $(k+r+l)$ coding group at the time of replication.
EC-Cache simply assigns slabs to coding groups comprising of random nodes.
Power-of-two-choices finds two candidate nodes at random for each slab, and picks the less loaded one.

\paragraph{\textbf{Probability of Data Loss Under Simultaneous Failures}}

\begin{figure}[t]
	\centering
	\subfloat[][Varied parities, $r$]{%
		\label{fig:analytic-r}%
		\includegraphics[width=0.5\columnwidth]{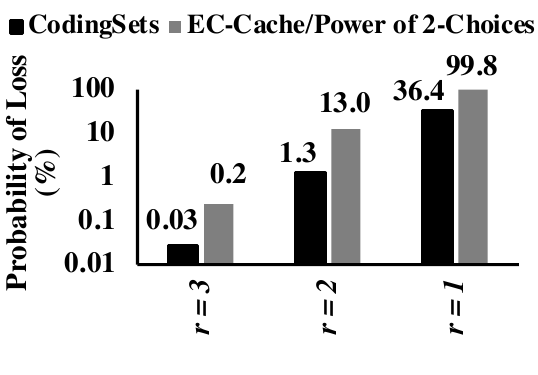}
	} 	
	\subfloat[][Varied load balancing factors, $l$]{%
		\label{fig:analytic-l}%
		\includegraphics[width=0.5\columnwidth]{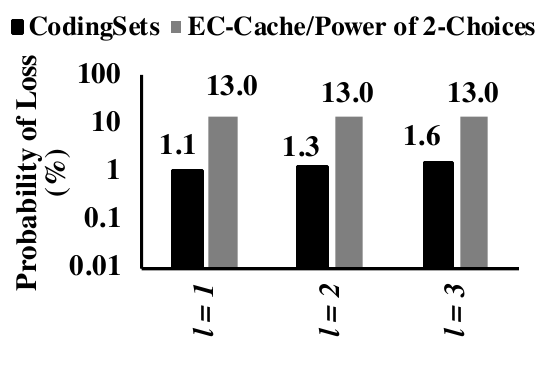}
	}\\\vspace{-3mm}
	\subfloat[][Varied slabs per machine, $S$]{%
	\label{fig:analytic-S}%
	\includegraphics[width=0.5\columnwidth]{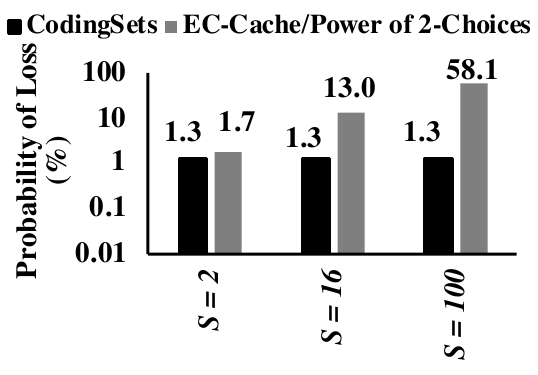}
	}
	\subfloat[][Varied failure rate, $f$]{%
		\label{fig:analytic-f}%
		\includegraphics[width=0.5\columnwidth]{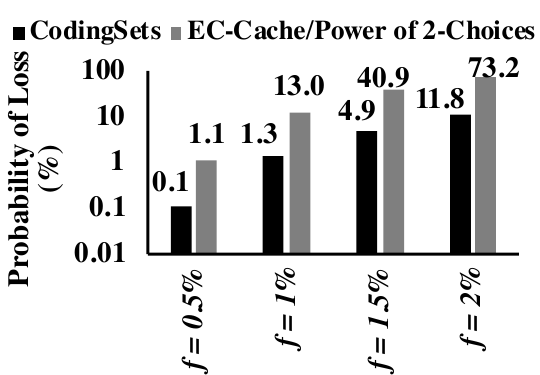}
	}
	\caption{Probability of data loss at different scenarios (base params. $k$=$8$, $r$=$2$, $l$=$2$, $S$=$16$, $f$=$1\%$) on a 1000-machine cluster. 
	}
	\label{fig:analytic}
\end{figure}

To evaluate the probability of data loss of {\name} under different scenarios in a large cluster setting, we compute
the probability of data loss under the three schemes.
Note that, in terms of data loss probability, we assume EC-Cache and power of two choices select random servers, and are therefore equivalent.
Figure~\ref{fig:analytic} compares the probabilities of loss for different parameters on a 1000-machine cluster.
Our baseline comparison is against the best case scenario for EC-Cache and power-of-two-choices, 
where the number of slabs per server is low (1~GB slabs, with 16~GB of memory per server). 

Even for a small number of slabs per server, {\name} reduces the probability of data loss by \emph{an order of magnitude}.
With a large number of slabs per server (\eg, 100) the probability of failure for EC-Cache becomes very high during correlated failure.
Figure~\ref{fig:analytic} shows that there is an inherent trade-off between the load-balancing factor ($l$) and the probability of data loss under correlated failures.

\begin{figure}[!t]
	\centering
	\includegraphics[scale=0.47]{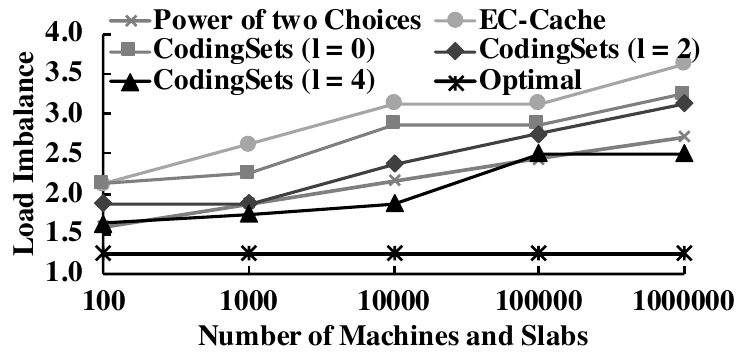}
	\caption{\copysets enhances {\name} with better load balancing across the cluster (base parameters k=8, r=2).}
	\label{fig:copysets-load-balance}
\end{figure}

\paragraph{\textbf{Load Balancing of \copysets}}
Figure~\ref{fig:copysets-load-balance} compares the load balancing of the three policies.
EC-Cache's random selection of $(k+r)$ nodes causes a higher load imbalance, since some nodes will randomly be overloaded more than others.
As a result, \copysets improves load balancing over EC-Cache scheme by $1.1\times$ even when $l = 0$,
since \copysets' coding groups are non-overlapping.
For $l = 4$, \copysets provides with $1.5\times$ better load balancing over EC-Cache at 1M machines.
The power of two choices improves load balancing by 0\%-20\% compared \copysets with $l=2$, because it has more degrees of freedom in choosing nodes,
but suffers from an order of magnitude higher failure rate (Figure~\ref{fig:analytic}).

\subsubsection{\textbf{Cluster Deployment}}
We run 250 containerized applications across 50 machines. 
For each application and workload, we create a container and randomly distribute it across the cluster. 
Here, total memory footprint is 2.76 TB; our cluster has 3.20 TB of total memory. 
Half of the containers use 100\% configuration; about 30\% use the 75\% configuration; and the rest use the 50\% configuration.
There are at most two simultaneous failures.

\begin{figure*}[t]
	\centering
	\subfloat[][{SSD Backup}]{%
		\label{fig:macro-is}%
		\includegraphics[scale=0.32]{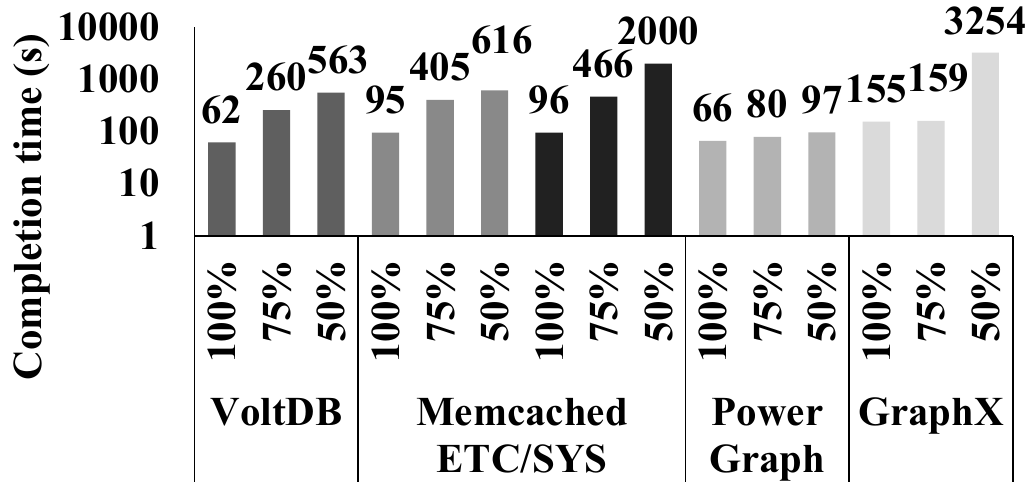}
	}
	\hfill
	\subfloat[][{\name}]{%
		\label{fig:macro-ec}%
		\includegraphics[scale=0.32]{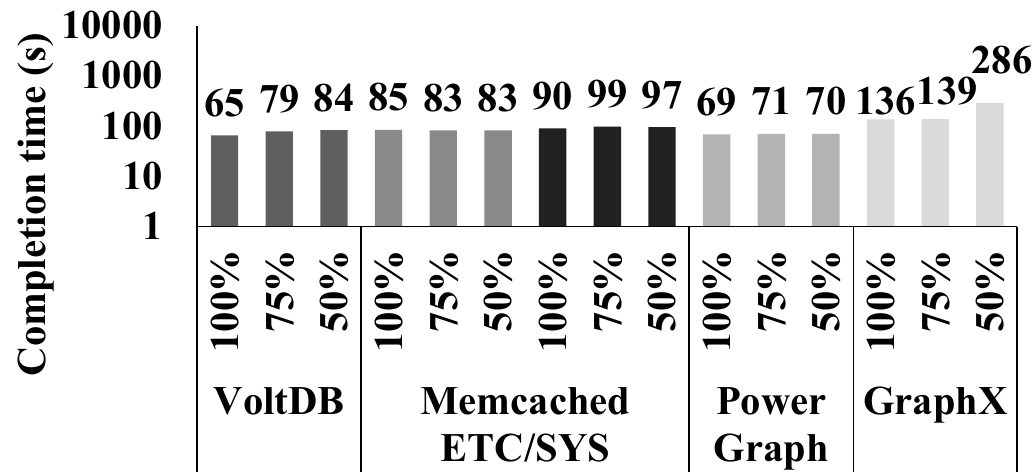}
	}
	\hfill
	\subfloat[][Replication]{%
		\label{fig:macro-rep}%
		\includegraphics[scale=0.32]{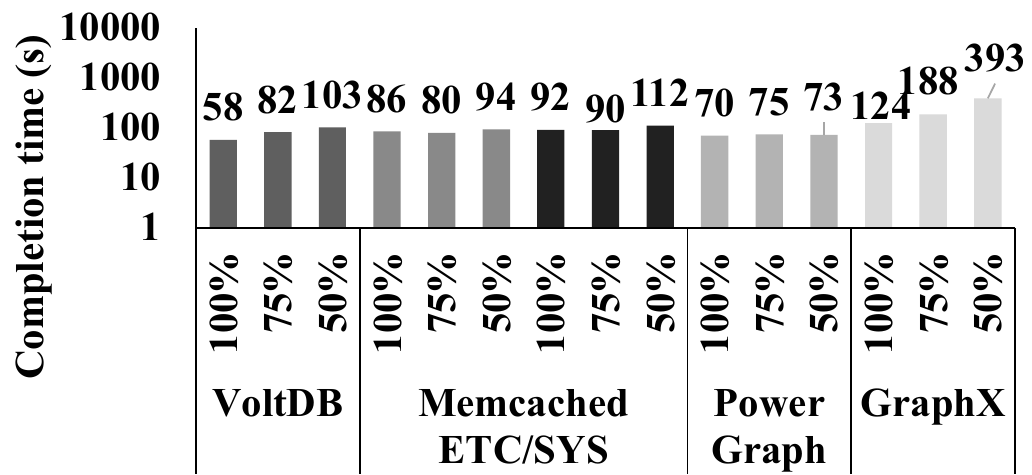}
	}
	\caption{Median completion times (\ie, throughput) of 250 containers on a 50-machine cluster.
	}
	\label{fig:macro-apps}	
\end{figure*}

\paragraph{\textbf{Application Performance}}
We compare application performance in terms of completion time (Figure~\ref{fig:macro-apps}) and latency (Table \ref{tab:macro-lat}) that demonstrate {\name}'s performance benefits in the presence of cluster dynamics. 
{\name}'s improvements increase with decreasing local memory ratio. 
Its throughput improvements w.r.t. SSD backup were up to 4.87$\times$ for 75\% and up to 20.61$\times$ for 50\%. 
Its latency improvements were up to 64.78$\times$ for 75\% and up to 51.47$\times$ for 50\%.
{\name}'s performance benefits are similar to replication (Figure~\ref{fig:macro-rep}), but with lower memory overhead.

\begin{table}[!t]
\vspace{2mm}
	\footnotesize
	\centering
	\begin{tabular}{|c|r|r|r|r||r|r|r|}            
		\hline
		\multicolumn{2}{|c|}{\multirow{2}{*}{Latency (ms)}} &
		\multicolumn{3}{c||}{ 50th} & \multicolumn{3}{c|}{ 99th}\\
		\multicolumn{2}{|c|}{}& SSD & HYD & REP & SSD & HYD & REP \\	
		\hline
		\multirow{3}{*}{VoltDB}	 
		& 100\% & 55  & 60 & 48 &179 & 173 & 177\\
		& 75\%	& 60  & 57 & 48 & 217 & 185 & 225 \\
		& 50\%  & 78  & 61 & 48 & 305 & 243 & 225\\
		\hline
		\multirow{3}{*}{\shortstack{ETC}}	 
		& 100\% & 138 & 119 & 118 & 260 & 245 & 247\\
		& 75\%  & 148 & 113 & 120 & 9912 & 240 & 263\\
		& 50\%  & 167 & 117 & 111 & 10175 & 244 & 259 \\
		\hline
		\multirow{3}{*}{\shortstack{SYS}}
		& 100\% & 145 & 127 & 125 & 249 & 269 & 267\\
		& 75\%  & 154 & 119 & 113 & 17557 & 271 & 321\\
		& 50\%  & 124 & 111 & 117 & 22828 & 452 & 356\\	 
		\hline
	\end{tabular}
	\caption{VoltDB and Memcached (ETC, SYS) latencies for SSD backup, {\name} (HYD) and replication (REP) in cluster setup.}
	\label{tab:macro-lat}
	\vspace{-4mm}
\end{table}

\begin{figure}[t]
	\centering
	\includegraphics[scale=0.45]{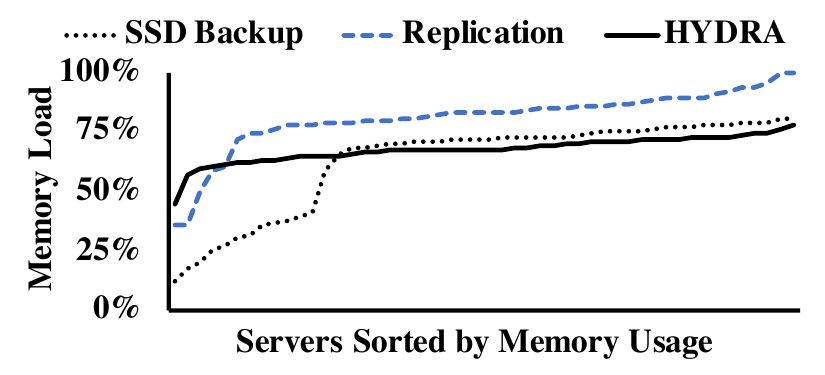}
	\caption{Average memory usage across 50 servers.	}
	\label{fig:macro-load}
\end{figure}

\paragraph{\textbf{Impact on Memory Imbalance and Stranding}}
Figure~\ref{fig:macro-load} shows that {\name} reduces memory usage imbalance w.r.t. coarser-grained memory management systems: in comparison to SSD backup-based (replication-based) systems, memory usage variation decreased from 18.5\% (12.9\%) to 5.9\% and the maximum-to-minimum utilization ratio decreased from 6.92$\times$ (2.77$\times$) to 1.74$\times$. 
{\name} better exploits unused memory in under-utilized machines, increasing the minimum memory utilization of any individual machine by 46\%. 
{\name} incurs about 5\% additional total memory usage compared to disk backup, whereas replication incurs 20\% overhead.

\subsection{Sensitivity Evaluation}
\label{sec:eval-sensitivity}

\begin{figure}[t]
	\centering    
	\subfloat[][r=4, $\Delta$=1]{%
		\label{fig:split}
		\begin{minipage}{0.16\textwidth}
			\includegraphics[width=\textwidth]{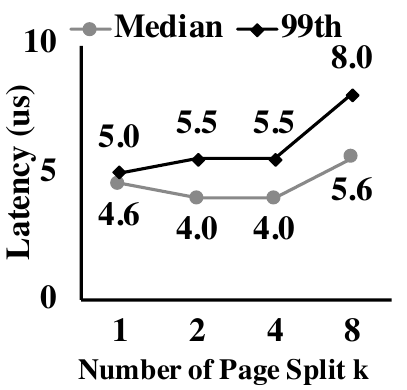}
		\end{minipage}
	}
	\subfloat[][k=8, r=4]{%
		\label{fig:delta}      
		\begin{minipage}{0.16\textwidth}
			\includegraphics[width=\textwidth]{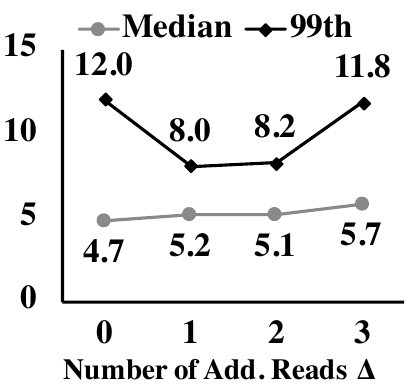}
		\end{minipage}
	}
	\subfloat[][k=8, $\Delta$=1]{
		\label{fig:parity}
		\begin{minipage}{0.15\textwidth}
			\includegraphics[width=\textwidth]{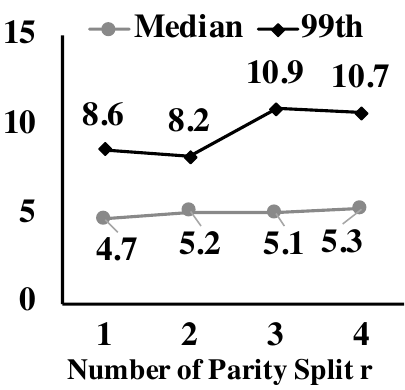}
			\end{minipage}
	}
	\caption{Impact of page splits ($k$), additional reads ($\Delta$) on read latency, and parity splits ($r$) on write latency.
	}
	\label{fig:paramter}
\end{figure}



\paragraph{\textbf{Impact of ($k$, $r$, $\Delta$) Choices}}
Figure~\ref{fig:split} shows read latency characteristics for varying $k$.
Increasing from $k$=1 to $k$=2 reduces median latency by parallelizing data transfers. 
Further increasing $k$ improves space efficiency (measured as $\frac{r}{k+r}$) and load balancing, but latency deteriorates as well.

Figure~\ref{fig:delta} shows read latency for varying values of $\Delta$. 
Although just one additional read (from $\Delta$=0 to $\Delta$=1) helps tail latency, more additional reads have diminishing returns; instead, it hurts latency due to proportionally increasing communication overheads. 
Figure~\ref{fig:parity} shows write latency variations for different $r$ values.
Increasing $r$ does not affect the median write latency. 
However, the tail latency increases from $r=3$ due to the increase in overall communication overheads.

\paragraph{\textbf{Resource Overhead}}
We measure average CPU utilization of {\name} components during remote I/O.
{\host} uses event-driven I/O and consumes only 0.001\% CPU cycles in each core.
Erasure coding causes 0.09\% extra CPU usage per core.
As {\name} uses one-sided RDMA, remote {\daemon}s do not have CPU overhead in the data path.

In cluster deployment, {\name} increases CPU utilization by 2.2\% on average and generates 291 Mbps RDMA traffic per machine, which is only 0.5\% of its 56 Gbps bandwidth.
Replication has negligible CPU usage but generates more than 1 Gbps traffic per machine.

\begin{table}[t]
	\vspace{2mm}
	\centering
	\small
	\begin{tabular}{|l|r|r|r|}
		\hline
		\textbf{Monthly Pricing} & Google & Amazon& Microsoft\\ 		
		\hline
		Standard machine   & \$1,553   & \$2,304 &  \$1,572\\
		1\% memory      	& \$5.18 & \$9.21  & \$5.92    \\
		\hline
		\hline
		\textbf{\name}      & 6.3\%    & 8.4\% & 7.3\%  \\
		\textbf{Replication}      & 3.3\%    & 4.8\% & 3.9\%  \\
		\textbf{PM Backup}     & 3.5\%    & 7.6\% & 4.9\%  \\
		\hline
	\end{tabular}
	\caption{Revenue model and TCO savings over three years for each machine with 30\% unused memory on average.}
	\label{tab:cost}
	\vspace{-4mm}
\end{table}

\begin{table*}[!t]
	\small
	\centering
	\begin{tabular}{c c c c c c }            
		\hline
		System  & Year & Deployability  &  Fault Tolerance & Load Balancing & {Latency Tolerance} \\ [0.5ex]         
		\hline\hline
		Memory Blade \cite{mem-ext}    				 & '09 & HW Change   & Reprovision  			   & None 			& None\\   
		RamCloud \cite{ramcloud-ft}       			 & '10 & App. Change & Remote Disks 			   & Power of Choices	& None\\
		FaRM \cite{farm}       						 & '14 & App. Change & Replication				   & Central Coordinator & None\\
		EC-Cache \cite{eccache}								 & '16 & App. Change & Erasure Coding 					   & Multiple Coding Groups & Late Binding \\
		{\is} \cite{infiniswap}     			     & '17 & Unmodified  & Local Disk 				   & Power of Choices & None\\
		Remote Regions \cite{remote-regions}     			     & '18 & App. Change  & None  				   & Central Manager & None\\		
		LegoOS \cite{legoos}     			     & '18 & OS Change  & Remote Disk 				   & None & None\\
		Compressed Far Memory \cite{far-memory} 	 & '19 & OS Change  & None 				   & None & None\\	
		Leap \cite{leap}     			     & '20 & OS Change  & None 				   & None & None\\	
		Kona \cite{kona}     			 & '21 & HW Change  & Replication 				   & None & None\\		
		{\name}                        				 &     & Unmodified  & Erasure Coding 			   & \copysets	  & Late Binding\\
		\hline
	\end{tabular}
	\caption{Selected proposals on remote memory in recent years.}
	\label{tab:related}
	\vspace{-4mm}
\end{table*}

\paragraph{\textbf{Background Slab Regeneration}}

To observe the overall latency to regenerate a slab, we manually evict one of the remote slabs. 
When it is evicted, {\host} places a new slab and provides the evicted slab information to the corresponding {\daemon}, which takes 54 ms. 
Then the {\daemon} randomly selects $k$ out of remaining remote slabs and read the page data, which takes 170 ms for a 1 GB slab. 
Finally, it decodes the page data to the local memory slab within 50 ms. 
Therefore, the total regeneration time for a 1 GB size slab is 274 ms, as opposed to taking several minutes to restart a server after failure.

To observe the impact of slab regeneration on disaggregated VMM, we run the  micro-benchmark mentioned in \S\ref{sec:eval-resilience}.
At the half-way of the application's runtime, we evict one of the remote slabs.
Background slab regeneration has a minimal impact on the remote read -- remote read latency increases by 1.09$\times$.
However, as remote writes to the victim slab halts until it gets regenerated, write latency increases by 1.31$\times$.

\subsection{TCO Savings}
\label{sec:eval-tco}
We limit our TCO analysis only to memory provisioning. 
The TCO savings of {\name} is the revenue from leveraged unused memory after deducting the TCO of RDMA hardware. 
We consider capital expenditure (CAPEX) of acquiring RDMA hardware and operational expenditure (OPEX) including their power usage over 3 years. 
An RDMA adapter costs \$600 \cite{mlx-adapter}, RDMA switch costs \$318 \cite{mlx-switch} per machine, and the operating cost is \$52 over 3 years \cite{infiniswap} -- overall, the 3-year TCO is \$970 for each machine. 
We consider the standard machine configuration and pricing from Google Cloud Compute \cite{google-price}, Amazon EC2 \cite{amazon-price}, and Microsoft Azure \cite{amazon-price} to build revenue models and calculate the TCO savings for 30\% of leveraged memory for each machine (Table \ref{tab:cost}). 
For example, in Google, the savings of disaggregation over 3 years using {\name} is ((\$5.18*30*36)/1.25-\$970)/(\$1553*36)*100\% = 6.3\%. 

\subsection{Disaggregation with Persistent Memory Backup}
\label{sec:pm-backup}
To observe the impact of persistent memory (PM), we run all the micro-benchmarks and real-world applications mentioned earlier over {\is} with local PM backup.
Unfortunately, at the time of writing, we cannot get hold of a real Intel Optane DC.
We emulate PM using DRAM with the latency characteristics mentioned in prior work~\cite{optane-dc-characteristics}.

Replacing SSD with local PM can significantly improve {\is}'s performance in a disaggregated cluster. 
However, for the micro-benchmark mentioned in \S\ref{sec:eval-resilience}, {\name} still provides 1.06$\times$ and 1.09$\times$ better 99th percentile latency over {\is} with PM backup during page-in and page-out, respectively.
Even for real-world applications mentioned in \S\ref{subsec:apps},  {\name} almost matches the performance of local PM backup -- application-level performance varies within 0.94--1.09$\times$ of that with PM backup.
Note that replacing SSD with PM throughout the cluster does not improve the availability guarantee in the presence of cluster-wide uncertainties.
Moreover, while resiliency through unused remote DRAM is free, PM backup costs $\$11.13$/GB~\cite{pm-price}.
In case of Google, the additional cost of \$2671.2 per machine for PM reduces the savings of disaggregation over 3 years from $6.3$\% to ((\$5.18*30*36)-\$970-\$2671.2)/(\$1553*36)*100\% = 3.5\% (Table \ref{tab:cost}).

%% file: related.tex
\section{Related Work}
\label{sec:related}

\paragraph{\textbf{Remote-Memory Systems}}
Many software systems tried leveraging remote machines' memory for paging~\cite{zahorjan-remote-paging, markatos-remote, nswap, transparent-remote-paging-vm, cashmere-vlm, swapping-infiniband, nbdx, mem-collab, infiniswap, far-memory, leap, kona}, global virtual memory abstraction \cite{global-memory-management, scalemp, para-remote}, and to create distributed data stores \cite{ramcloud-ft, farm, mica, rocksteady, aurora, polardb, fasst, AIFM}.
Hardware-based remote access to memory using PCIe interconnects \cite{mem-ext} and extended NUMA fabric \cite{soNUMA} are also proposed.
Table~\ref{tab:related} compares a selected few.

\paragraph{\textbf{Cluster Memory Solutions} }
With the advent of RDMA, there has been a renewed interest in cluster memory solutions.
The primary way of leveraging cluster memory is through key-value interfaces \cite{mitchell2013using, kaminsky2014using, farm, ramcloud-ft}, distributed shared memory \cite{power2010building, nelson2015latency}, or distributed lock \cite{yoon2018distributed}. However, these solutions are either limited by their interface or replication overheads. {\name}, on the contrary, is a transparent, memory-efficient, and load-balanced mechanism for resilient remote memory.

\paragraph{\textbf{Erasure Coding in Storage}}
Erasure coding has been widely employed in RAID systems to achieve space-efficient fault tolerance \cite{datacenter-ec, XORingEN}. 
Recent large-scale clusters leverage erasure coding for storing {\em cold} data in a space-efficient manner to achieve fault-tolerance \cite{ceph, ec-azure, facebook-blob}. 
EC-Cache \cite{eccache} is an erasure-coded in-memory cache for 1MB or larger objects, but it is highly susceptible to data loss under correlated failures, and its scalability is limited due to communication overhead. 
In contrast, {\name} achieves resilient erasure-coded remote memory with single-digit $\mu$s page access latency.

%% file: outro.tex
\section{Conclusion}
\label{sec:outro}
%
{\name} leverages online erasure coding to achieve single-digit $\mu$s latency under failures, while judiciously placing erasure-coded data using \copysets to improve availability and load balancing.
It matches the resilience of replication with 1.6$\times$ lower memory overhead and significantly improves latency and throughput of real-world memory-intensive applications over SSD backup-based resilience. 
Furthermore, \copysets allows {\name} to reduce the probability of data loss under simultaneous failures by about 10$\times$.
Overall, {\name} makes resilient remote memory practical.

%% file: acknowledgments.tex
\section*{Acknowledgments}
\label{sec:acknowledgments}
\rev{
We thank the anonymous reviewers, our shepherd, 
Danyang Zhuo, and SymbioticLab members for their insightful feedback 
that helped improve the paper. 
This work was supported in part by National Science Foundation grants (CNS-1845853, CNS-1900665, CNS-2104243) and a gift from VMware.
}